\documentclass[12pt,doublespacing]{article}

\usepackage{fullpage}
\usepackage{setspace}
\usepackage{authblk}

\usepackage{cite}
\usepackage{hyperref} 

\usepackage{graphicx}

\usepackage{amsmath}
\usepackage{amssymb}

\usepackage{float}
\usepackage{subfigure}
\usepackage{makecell}
\usepackage{rotating}

\usepackage{enumitem}

\begin{document}

\title{Analysis of Electromagnetic Scattering from Semiconductor Nanostructures by Solving Coupled Volume Integral and Two-fluid Hydrodynamic Equations}

\author[1]{Doolos Aibek Uulu}
\author[2]{Meruyert Khamitova}
\author[3]{Rui Chen}
\author[2]{Liang Chen}
\author[4]{Ping Li}
\author[2]{Hakan Bagci\vspace{0.5cm}}

\affil[1]{Department of Cyber Security, Light Academy College of Engineering, 
\authorcr Bishkek, Kyrgyzstan\vspace{0.25cm}}

\affil[2]{Electrical and Computer Engineering Program,
\authorcr Computer, Electrical, and Mathematical Sciences and Engineering Division,
\authorcr King Abdullah University of Science and Technology (KAUST), 
\authorcr Thuwal, Saudi Arabia\vspace{0.25cm}}

\affil[3]{School of Microelectronics, Nanjing University of Science and Technology (NJUST),
\authorcr Nanjing, China\vspace{0.25cm}}

\affil[4]{School of Electronic Science and Engineering, University of Electronic Science and Technology of China (UESTC), 
\authorcr Chengdu, China\vspace{0.25cm}}

%\email{\authormark{*}opex@optica.org} %% email address is required


\date{}
\maketitle
\newpage

\begin{abstract}
Semiconductor-based plasmonic nanostructures support localized surface plasmon modes in the infrared region. Unlike metallic nanostructures, they support both free electrons and holes, requiring a two-fluid hydrodynamic Drude equation (HDE) to accurately capture spatial dispersion effects and low-frequency acoustic plasmon modes that cannot be described by single-fluid models. In this work, a volume integral equation (VIE)-based solver is proposed for the analysis of electromagnetic scattering from semiconductor nanostructures. The proposed approach couples the VIE, formulated in terms of the electric flux density and the free-electron and hole polarization currents, with the two-fluid HDE. The coupled system is discretized using a tetrahedral mesh and solved efficiently using a two-level iterative solver. In contrast to finite-element-based methods, the proposed VIE-based approach does not require domain-wide meshing and inherently satisfies the radiation condition, thereby eliminating artificial absorbing boundaries. Numerical results for InSb-type semiconductor nanostructures demonstrate the accuracy and efficiency of the proposed VIE-based solver and its ability to capture unique optical phenomena, such as acoustic plasmon resonances and the blueshift of localized surface plasmon resonances, that cannot be described by the single-fluid HDE or classical Drude-based models.
\end{abstract}
\newpage

%%%%%%%%%%%%%%%%%%%%%%%%%%  body  %%%%%%%%%%%%%%%%%%%%%%%%%%
\section{Introduction}
In recent decades, plasmonic nanostructures have attracted significant attention across various scientific and engineering disciplines due to their ability to confine electromagnetic fields to subwavelength regions and their strong, geometry-dependent optical response~\cite{naik2013alternative, murray2007plasmonic, stockman2011nanoplasmonics}. These characteristics have enabled numerous practical applications, including near-field nanoimaging systems~\cite{kawata2009plasmonics}, superlenses~\cite{li2018cascaded}, nanoantennas~\cite{ross2011omnidirectional, fischer2008engineering}, resonators~\cite{kewes2018heuristic, wang2009transmission}, waveguides~\cite{kazanskiy2020plasmonic}, couplers~\cite{uulu2019fourier, veronis2007theoretical}, and sensors~\cite{homola1999surface,lin2008surface, khonina2021plasmonic}.

Plasmonic nanostructures are typically realized using metals because of their ability to support localized surface plasmon (LSP) modes at optical frequencies~\cite{chang2005surface}. At these frequencies, the dielectric permittivity of a metal is commonly described by the classical Drude model~\cite{maier2007plasmonics}. However, as the structural dimensions approach the nanoscale, this model fails to predict the experimentally observed blueshift in plasmon resonance frequencies~\cite{tiggesbaumker1993blue}. This shift is a direct consequence of the quantum behavior of free electrons, which gives rise to spatial dispersion, a nonlocal response in which the electric field at a given point depends not only on the local excitation but also on the fields in its vicinity. Although full quantum mechanical models, such as time-dependent density functional theory~\cite{marques2006time}, can accurately capture these interactions, their high computational cost limits their applicability to structures only a few nanometers in size. A more computationally tractable alternative is the hydrodynamic Drude equation (HDE), a quasi-classical approach that treats the electron plasma as an electron fluid and effectively captures the spatial dispersion effects responsible for the blueshift~\cite{forstmann1986metal}.

More recently, semiconductors~\cite{tang2020plasmonic,barho2016all, li2011all, berrier2010ultrafast} have emerged as alternative plasmonic materials. Unlike metals, which support LSP modes at optical frequencies, semiconductors exhibit these modes in the infrared region, thereby extending the range of accessible applications. Furthermore, their electrical properties can be tuned via doping~\cite{zhou2015comparative} or external biasing~\cite{liu2017electrical}, offering an additional 
degree of control over the plasmonic response. Since semiconductors operate in the infrared regime, spatial dispersion effects become significant at larger structural dimensions than in metals, which relaxes the fabrication requirements for plasmonic nanostructures~\cite{deceglia2018viscoelastic}. 

Unlike metals, in which free electrons are the only charge carriers, semiconductors support both free electrons and holes, leading to the coexistence of free electron and hole plasmas~\cite{munekata1986densities}. The conventional single-fluid HDE therefore cannot adequately describe the carrier dynamics in semiconductors. To address this, a two-fluid HDE has been proposed~\cite{maack2018two}, in which the motions of free electrons and holes are modeled separately by two single-fluid HDEs with distinct parameters, coupled through the electric field. This model captures unique optical properties of semiconductor nanostructures, such as low-frequency acoustic plasmon resonances~\cite{dong2020modified}, which cannot be described by the single-fluid HDE. 

The polarization currents associated with free electrons and holes generate electromagnetic fields inside the semiconductor, and their interaction can be described by a coupled system of Maxwell equations and the two-fluid HDE. Due to the complexity of this coupled system and the need to handle arbitrarily shaped nanostructure geometries, numerical methods are required for its solution. A finite element method (FEM) has been developed for the analysis of semiconductor nanowires and nanodimers~\cite{golestanizadeh2019hydrodynamic}. However, like conventional FEM solvers for Maxwell equations, this approach requires discretization of the entire computation domain and the use of approximate absorbing boundary conditions~\cite{jin2015}, both of which limit the accuracy and efficiency of the solver.

To overcome these shortcomings, a volume integral equation (VIE)-based solver is proposed in this work to analyze electromagnetic scattering from semiconductor nanostructures. Within the VIE framework, only the nanostructure is discretized, and the radiation condition is inherently satisfied through the scalar Green function of the background medium. The proposed approach couples the VIE with the two-fluid HDE. The VIE expresses the electric field as a volumetric convolution of the scalar Green function with the electric flux density and the free-electron and hole polarization currents, while the two-fluid HDE relates the electric field to these polarization currents. To numerically solve the coupled system, the nanostructure is discretized using a tetrahedral mesh, and the unknown electric flux density and polarization currents are expanded using Schaubert-Wilton-Glisson (SWG) basis functions~\cite{schaubert1984tetrahedral} defined on the mesh. Substituting these expansions into the coupled VIE and two-fluid HDE and applying Galerkin testing yield a matrix system for the unknown expansion coefficients, which is solved efficiently using a two-level iterative solver~\cite{aibek2022solution}. Numerical results demonstrate the accuracy and efficiency of the proposed VIE-based solver and its ability to capture unique optical phenomena in semiconductor nanostructures, such as acoustic plasmon resonances, that cannot be described by the single-fluid HDE or classical Drude-based models.

\section{Formulation}
\subsection{Coupled System of VIE and Two-fluid HDE}
Let $V$ denote the region occupied by a semiconductor nanostructure embedded in an unbounded homogeneous background medium with permittivity $\varepsilon_0$ and permeability $\mu_0$. The structure is excited by an incident electric field $\mathbf{E}^\mathrm{inc}(\mathbf{r})$,  where a time-harmonic dependence of the form $e^{\mathrm{j}\omega t}$  is assumed, and the formulation is developed in the frequency domain. Here, $\omega$ is the angular frequency and $t$ denotes time. The electromagnetic response within $V$ consists of contributions from bound electrons, free electrons, and holes. The bound-electron response is characterized by the relative permittivity $\varepsilon_\mathrm{b}(\mathbf{r})$, while the free-electron and hole contributions are characterized by polarization currents induced within $V$.  Accordingly, the total polarization current is expressed as
\begin{equation}
    \mathbf{J}(\mathbf{r}) = \mathbf{J}_{\mathrm{e}}(\mathbf{r})+\mathbf{J}_{\mathrm{h}}(\mathbf{r})+\mathrm{j} \omega \varepsilon_{0} (\varepsilon_{\mathrm{b}}(\mathbf{r})-1) \mathbf{E}(\mathbf{r})
    \label{eq:total_cur}
\end{equation}
where $\mathbf{J}_{\mathrm{e}}(\mathbf{r})$ and $\mathbf{J}_{\mathrm{h}}(\mathbf{r})$ denote the polarization currents associated with free electrons and holes, respectively, while the last term represents the bound-electron contribution in terms of the total electric field $\mathbf{E}(\mathbf{r})$. This total field is decomposed into the incident field $\mathbf{E}^{\mathrm{inc}}(\mathbf{r})$ and the scattered field $\mathbf{E}^{\mathrm{sca}}(\mathbf{r})$ as
\begin{equation}
\mathbf{E}(\mathbf{r})=\mathbf{E}^{\mathrm{inc}}(\mathbf{r})+\mathbf{E}^{\mathrm{sca}}(\mathbf{r}).    
\label{eq:total_field}
\end{equation}
The scattered electric field $\mathbf{E}^{\mathrm{sca}}(\mathbf{r})$ generated by the induced current $\mathbf{J}(\mathbf{r})$ is expressed using the volume integral operator as
\begin{equation}
    \mathbf{E}^{\mathrm{sca}}(\mathbf{r})=\mathcal{L}[\mathbf{J}](\mathbf{r})
    \label{eq:e_sca}
\end{equation}
where the operator $\mathcal{L}[\mathbf{X}](\mathbf{r})$ is defined as
\begin{equation}
\mathcal{L}[\mathbf{X}](\mathbf{r})=-\mathrm{j} \omega \mu_{0} \int_{V} \mathbf{X}(\mathbf{r}^{\prime}) G(\mathbf{r}, \mathbf{r}^{\prime})\,d v^{\prime}+\frac{1}{\mathrm{j} \omega \varepsilon_{0}}\nabla \nabla \cdot \int_{V} \mathbf{X}(\mathbf{r}^{\prime}) G(\mathbf{r}, \mathbf{r}^{\prime})\, d v^{\prime}.
\label{eq:integral_operator}
\end{equation}
Here, $G(\mathbf{r}, \mathbf{r}^{\prime})=\exp (-\mathrm{j} k_{0} \left|\mathbf{r}-\mathbf{r}^{\prime}\right|) /(4 \pi \left|\mathbf{r}-\mathbf{r}^{\prime}\right|)$ is the scalar Green function of the homogeneous background medium, $k_{0}=\omega \sqrt{\mu_{0} \varepsilon_{0}}$ is the background medium wavenumber, and the gradient operators in~\eqref{eq:integral_operator} act on the observation variable $\mathbf{r}$. 

To eliminate $\mathbf{E}(\mathbf{r})$ as an explicit unknown, the electric flux density $\mathbf{D}(\mathbf{r})$ is defined as
\begin{equation}
\label{eq:flux}
\mathbf{D}(\mathbf{r}) = \varepsilon_0 \mathbf{E}(\mathbf{r})+\frac{1}{\mathrm{j}\omega}\mathbf{J}(\mathbf{r}).
\end{equation}
This choice is motivated by the fact that the normal component of $\mathbf{D}(\mathbf{r})$ is continuous across material interfaces within $V$. Substituting Eq.~\eqref{eq:flux} into Eq.~\eqref{eq:total_cur} to eliminate $\mathbf{E}(\mathbf{r})$ yields
\begin{equation}
\label{eq:total_cur2}
\mathbf{J}(\mathbf{r})=\mathrm{j} \omega \kappa(\mathbf{r}) \mathbf{D}(\mathbf{r})+\frac{\mathbf{J}_{\mathrm{e}}(\mathbf{r})}{\varepsilon_{\mathrm{b}}(\mathbf{r})}+\frac{\mathbf{J}_{\mathrm{h}}(\mathbf{r})}{\varepsilon_{\mathrm{b}}(\mathbf{r})}
\end{equation}
where $\kappa(\mathbf{r}) = 1-1/{\varepsilon _{\mathrm{b}}(\mathbf{r})}$. Substituting Eq.~\eqref{eq:total_cur} into Eq.~\eqref{eq:flux} and solving for $\mathbf{E}(\mathbf{r})$ gives
\begin{equation}
\mathbf{E}(\mathbf{r})=\frac{\mathbf{D}(\mathbf{r})}{\varepsilon_0 \varepsilon_{\mathrm{b}}(\mathbf{r})}-\frac{\mathbf{J}_{\mathrm{e}}(\mathbf{r})}{\mathrm{j}\omega \varepsilon_0 \varepsilon_{\mathrm{b}}(\mathbf{r})}-\frac{\mathbf{J}_{\mathrm{h}}(\mathbf{r})}{\mathrm{j}\omega\varepsilon_0 \varepsilon_{\mathrm{b}}(\mathbf{r})}.
\label{eq:total_field2}
\end{equation}
Substituting Eq.~\eqref{eq:total_cur2} into Eq.~\eqref{eq:e_sca}, and using the resulting expression together with Eq.~\eqref{eq:total_field2} in Eq.~\eqref{eq:total_field} for $\mathbf{r} \in V$ yields the VIE in unknowns $\mathbf{D}(\mathbf{r})$, $\mathbf{J}_{\mathrm{e}}(\mathbf{r})$, and $\mathbf{J}_{\mathrm{h}}(\mathbf{r})$ as
\begin{equation}
\label{eq:vie}
\begin{aligned}
\mathbf{E}^{\mathrm{inc}}(\mathbf{r}) &=\frac{\mathbf{D}(\mathbf{r})}{\varepsilon_{0} \varepsilon_{\mathrm{b}}(\mathbf{r})}-\mathrm{j} \omega \mathcal{L}[\kappa \mathbf{D}](\mathbf{r}) \\
&-\frac{\mathbf{J}_{\mathrm{e}}(\mathbf{r})}{\mathrm{j} \omega \varepsilon_{0} \varepsilon_{\mathrm{b}}(\mathbf{r})}-\mathcal{L}[\frac{\mathbf{J}_{\mathrm{e}}}{\varepsilon_{\mathrm{b}}}](\mathbf{r}) -\frac{\mathbf{J}_{\mathrm{h}}(\mathbf{r})}{\mathrm{j} \omega \varepsilon_{0} \varepsilon_{\mathrm{b}}(\mathbf{r})}-\mathcal{L}[\frac{\mathbf{J}_{\mathrm{h}}}{\varepsilon_{\mathrm{b}}}](\mathbf{r}),\, \mathbf{r} \in V. 
\end{aligned}
\end{equation}

The dynamics of free electrons and holes within $V$ are governed by the two-fluid HDE~\cite{maack2018two}:
\begin{equation}
\label{eq:hydro}
\begin{aligned}
\beta_{\mathrm{e}}^{2} \nabla\left[\nabla \cdot \mathbf{J}_{\mathrm{e}}(\mathbf{r})\right]+\omega(\omega-\mathrm{j} \gamma_{\mathrm{e}}) \mathbf{J}_{\mathrm{e}}(\mathbf{r})&=-\mathrm{j} \omega \omega_{\mathrm{e}}^{2} \varepsilon_{0} \mathbf{E}(\mathbf{r})\\
\beta_{\mathrm{h}}^{2} \nabla\left[\nabla \cdot \mathbf{J}_{\mathrm{h}}(\mathbf{r})\right]+\omega(\omega-\mathrm{j} \gamma_{\mathrm{h}}) \mathbf{J}_{\mathrm{h}}(\mathbf{r}) &=-\mathrm{j} \omega \omega_{\mathrm{h}}^{2} \varepsilon_{0} \mathbf{E}(\mathbf{r}), \, \mathbf{r} \in V.
\end{aligned}
\end{equation}
Here, $\omega_\mathrm{a}$, $\beta_\mathrm{a}$, and $\gamma_\mathrm{a}$, $\mathrm{a} \in \{\mathrm{e},\mathrm{h} \}$, denote the plasma frequencies, the hydrodynamic pressure parameters, and the damping constants, respectively, for the free electrons ($\mathrm{a} = \mathrm{e}$) and the holes ($\mathrm{a} = \mathrm{h}$).  The pressure terms $\beta^2_{\mathrm{e}}\nabla [\nabla\cdot \mathbf{J}_{\mathrm{e}}(\mathbf{r})]$ and $\beta^2_{\mathrm{h}}\nabla [\nabla\cdot \mathbf{J}_{\mathrm{h}}(\mathbf{r})]$ in Eq.~\eqref{eq:hydro} give rise to longitudinal electric-field components inside the semiconductor~\cite{golestanizadeh2019hydrodynamic}, in addition to the transverse components described by local Drude-based models, which necessitates two additional boundary conditions. Specifically, the normal components of $\mathbf{J}_{\mathrm{e}}(\mathbf{r})$ and $\mathbf{J}_{\mathrm{h}}(\mathbf{r})$ are required to vanish on the boundary surface $S$ of $V$:
\begin{equation}
\label{eq:add_BC}
\begin{aligned}
\hat{\mathbf{n}}(\mathbf{r}) \cdot \mathbf{J}_{\mathrm{e}}(\mathbf{r})&=0\\
\hat{\mathbf{n}}(\mathbf{r}) \cdot \mathbf{J}_{\mathrm{h}}(\mathbf{r})&=0
\end{aligned}
\end{equation}
where $\hat{\mathbf{n}}(\mathbf{r})$ is the outward-pointing unit normal vector at $\mathbf{r} \in S$. Physically, these boundary conditions imply that neither free electrons nor holes can flow across the semiconductor boundary into the surrounding background medium.

Substituting Eq.~\eqref{eq:total_field2} into Eq.~\eqref{eq:hydro} to eliminate $\mathbf{E}(\mathbf{r})$ yields the two-fluid HDE in unknowns $\mathbf{D}(\mathbf{r})$, $\mathbf{J}_{\mathrm{e}}(\mathbf{r})$, and $\mathbf{J}_{\mathrm{h}}(\mathbf{r})$ as
\begin{equation}
\label{eq:hydro_coupled}
\begin{aligned}
\beta_{\mathrm{e}}^{2} \nabla[\nabla \cdot \mathbf{J}_{\mathrm{e}}(\mathbf{r})]+\Big(\omega(\omega-\mathrm{j} \gamma_{\mathrm{e}})-\frac{\omega_{\mathrm{e}}^{2}}{\varepsilon_{\mathrm{b}}(\mathbf{r})}\Big) \mathbf{J}_{\mathrm{e}}(\mathbf{r})-\frac{\omega_{\mathrm{e}}^{2}}{\varepsilon_{\mathrm{b}}(\mathbf{r})}\mathbf{J}_{\mathrm{h}}(\mathbf{r})&=-\mathrm{j} \omega \omega_{\mathrm{e}}^{2} \frac{\mathbf{D}(\mathbf{r})}{\varepsilon_{\mathrm{b}}(\mathbf{r})}\\
\beta_{\mathrm{h}}^{2} \nabla[\nabla \cdot \mathbf{J}_{\mathrm{h}}(\mathbf{r})]+\Big(\omega(\omega-\mathrm{j} \gamma_{\mathrm{h}})-\frac{\omega_{\mathrm{h}}^{2}}{\varepsilon_{\mathrm{b}}(\mathbf{r})}\Big) \mathbf{J}_{\mathrm{h}}(\mathbf{r})-\frac{\omega_{\mathrm{h}}^{2}}{\varepsilon_{\mathrm{b}}(\mathbf{r})}\mathbf{J}_{\mathrm{e}}(\mathbf{r})&=-\mathrm{j} \omega \omega_{\mathrm{h}}^{2} \frac{\mathbf{D}(\mathbf{r})}{\varepsilon_{\mathrm{b}}(\mathbf{r})},\, \mathbf{r} \in V.
\end{aligned}
\end{equation}
Equations~\eqref{eq:vie} and~\eqref{eq:hydro_coupled} together constitute the coupled system of the VIE and the two-fluid HDE in the three unknowns $\mathbf{D}(\mathbf{r})$, $\mathbf{J}_{\mathrm{e}}(\mathbf{r})$, and $\mathbf{J}_{\mathrm{h}}(\mathbf{r})$.

\subsection{Discretization}\label{sec:disc}
To solve the coupled system of VIE and the two-fluid HDE, Eqs.~\eqref{eq:vie} and~\eqref{eq:hydro_coupled}, the region $V$ is first discretized using a tetrahedral mesh. The unknowns $\mathbf{D}(\mathbf{r})$, $\mathbf{J}_{\mathrm{e}}(\mathbf{r})$, and $\mathbf{J}_{\mathrm{h}}(\mathbf{r})$ are then expanded using the SWG basis functions~\cite{schaubert1984tetrahedral} as
\begin{equation}
\label{eq:basis_expansion}
\begin{aligned}
\mathbf{D}(\mathbf{r})&=\sum_{n=1}^{N_{\mathrm{D}}}\{\bar{I}_{\mathrm{D}}\}_{n} \mathbf{f}^{\mathrm{D}}_n(\mathbf{r})\\
\mathbf{J}_{\mathrm{e}}(\mathbf{r})&=\sum_{n=1}^{N_{\mathrm{e}}}\{\bar{I}_{\mathrm{e}}\}_{n} \mathbf{f}_{n}^{\mathrm{e}}(\mathbf{r})\\
\mathbf{J}_{\mathrm{h}}(\mathbf{r})&=\sum_{n=1}^{N_{\mathrm{h}}}\{\bar{I}_{\mathrm{h}}\}_{n} \mathbf{f}^{\mathrm{h}}_n(\mathbf{r}). 
\end{aligned}
\end{equation}
Here, $\mathbf{f}_{n}^{\mathrm{a}}(\mathbf{r})$, with $\mathrm{a} \in \{\mathrm{D},\mathrm{e},\mathrm{h} \}$, denote the SWG basis functions used to expand $\mathbf{D}(\mathbf{r})$, $\mathbf{J}_{\mathrm{e}}(\mathbf{r})$, and $\mathbf{J}_{\mathrm{h}}(\mathbf{r})$, respectively, $N_\mathrm{a}$ is the corresponding number of basis functions, and $\{\bar{I}_{\mathrm{a}}\}_{n}$ are the unknown expansion coefficients. An SWG basis function $\mathbf{f}_{n}(\mathbf{r})$, $\mathbf{r} \in V_n$, is defined on the triangular face $S_n$ shared by a pair of adjacent tetrahedrons as~\cite{schaubert1984tetrahedral}
\begin{align}
\mathbf{f}_{n}(\mathbf{r})=\left\{\begin{array}{l}
\mathbf{f}_{n}^{+}(\mathbf{r})=\frac{A_{n}}{3 \Omega_{n}^{+}}
(\mathbf{r}-\mathbf{r}_{n}^{+}), \quad \mathbf{r} \in V_{n}^{+} \\
\mathbf{f}_{n}^{-}(\mathbf{r})=-\frac{A_{n}}{3 \Omega_{n}^{-}}
(\mathbf{r}-\mathbf{r}_{n}^{-}), \quad \mathbf{r} \in V_{n}^{-}
\end{array}.\right.
\label{eq:swg_definition}
\end{align}
Here, $A_n$ denotes the area of $S_n$, $V_n^+$ and $V_n^-$ are the adjacent tetrahedrons sharing $S_n$, $\mathbf{r}_n^{\pm}$ are the free vertices of $V_n^{\pm}$ that do not lie on $S_n$, $\Omega_n^{\pm}$ denote the volumes of $V_n^{\pm}$, and $V_n = V_n^{+} \cup V_n^{-}$.

Since the normal component of $\mathbf{D}(\mathbf{r})$ is continuous across $S$, the basis functions $\mathbf{f}^{\mathrm{D}}_n(\mathbf{r})$ include ``full'' SWG basis functions  $\mathbf{f}_{n}(\mathbf{r})$ defined on every pair of tetrahedrons inside $V$ and ``half'' SWG basis functions $\mathbf{f}_{n}^{+}(\mathbf{r})$ defined on one tetrahedron having a triangular face on $S$. In contrast, because the normal components of $\mathbf{J}_{\mathrm{e}}(\mathbf{r})$ and $\mathbf{J}_{\mathrm{h}}(\mathbf{r})$ vanish on $S$, as required by Eq.~\eqref{eq:add_BC}, the basis functions $\mathbf{f}^{\mathrm{e}}_n(\mathbf{r})$ and $\mathbf{f}^{\mathrm{h}}_n(\mathbf{r})$ include only ``full'' SWG basis functions $\mathbf{f}_{n}(\mathbf{r})$ defined on every pair of tetrahedrons inside $V$. This choice of basis functions naturally enforces the boundary conditions in Eq.~\eqref{eq:add_BC}. 

Finally, substituting the expansions in Eq.~\eqref{eq:basis_expansion} into Eqs.~\eqref{eq:vie} and~\eqref{eq:hydro_coupled} and applying Galerkin testing with $\mathbf{f}_{m}^{\mathrm{a}}(\mathbf{r})$, ${m}=1,2,\dots,N_\mathrm{a}$, and $\mathrm{a} \in \{\mathrm{D},\mathrm{e},\mathrm{h} \}$, yield a matrix system of size $(N_{\mathrm{D}}+N_{\mathrm{e}}+N_{\mathrm{h}}) \times(N_{\mathrm{D}}+N_{\mathrm{e}}+N_{\mathrm{h}})$:
\begin{equation}
\underbrace{\begin{bmatrix}
\bar{\bar{Z}}_{\mathrm{DD}} & \bar{\bar{Z}}_{\mathrm{De}} &\bar{\bar{Z}}_{\mathrm{Dh}}\\
\bar{\bar{Z}}_{\mathrm{eD}} & \bar{\bar{Z}}_{\mathrm{ee}}& \bar{\bar{Z}}_{\mathrm{eh}}\\
\bar{\bar{Z}}_{\mathrm{hD}} & \bar{\bar{Z}}_{\mathrm{he}}& \bar{\bar{Z}}_{\mathrm{hh}}
\end{bmatrix}}_{\bar{\bar{Z}}} \underbrace{\begin{bmatrix}
\bar{I}_{\mathrm{D}} \\
\bar{I}_{\mathrm{e}} \\
\bar{I}_{\mathrm{h}}
\end{bmatrix}}_{\bar{I}}=\underbrace{\begin{bmatrix}
\bar{V}^{\mathrm{inc}} \\
0\\
0
\end{bmatrix}}_{\bar{V}} .
\label{eq:matrix_system}
\end{equation}
Here, $\bar{V}^{\mathrm{inc}}$ is the excitation vector, whose entries contain the tested incident field values, and the zeros in the second and third rows of the right-hand side vector follow from the fact that the two-fluid HDE has no excitation source term. Since $\mathcal{L}[\cdot](\cdot)$ is a global operator due to the Green function kernel, only $\bar{\bar{Z}}_{\mathrm{DD}}$, $\bar{\bar{Z}}_{\mathrm{De}}$, and $\bar{\bar{Z}}_{\mathrm{Dh}}$ are dense, while the remaining blocks are sparse, each having at most seven nonzero entries per row. The expressions of the matrix entries are provided in the Appendix.

\subsection{Solution of the Matrix System}\label{sol:mat}
To solve the matrix system in Eq.~\eqref{eq:matrix_system}, a two-level iterative solver~\cite{aibek2022solution} is employed. For this purpose, Eq.~\eqref{eq:matrix_system} is first rewritten as
\begin{align}
\underbrace{\begin{bmatrix}
\bar{\bar{Z}}_{\mathrm{DD}} & \bar{\bar{Z}}_{\mathrm{Dy}} \\
\bar{\bar{Z}}_{\mathrm{yD}} & \bar{\bar{Z}}_{\mathrm{yy}}
\end{bmatrix}}_{\bar{\bar{Z}}} \underbrace{\begin{bmatrix}
\bar{I}_{\mathrm{D}} \\
\bar{I}_{\mathrm{y}}
\end{bmatrix}}_{\bar{I}}=\underbrace{\begin{bmatrix}
\bar{V}^{\mathrm{inc}} \\
0
\end{bmatrix}}_{\bar{V}} 
\label{eq:matrix_system2}
\end{align}
where $\bar{I}_{\mathrm{y}}=[\bar{I}_{\mathrm{e}}\;\bar{I}_{\mathrm{h}}]^{\mathrm{T}}$, $\bar{\bar{Z}}_{\mathrm{Dy}}=[\bar{\bar{Z}}_{\mathrm{De}}\;\bar{\bar{Z}}_{\mathrm{Dh}}]$ of size $N_{\mathrm{D}} \times (N_{\mathrm{e}}+N_{\mathrm{h}})$, $\bar{\bar{Z}}_{\mathrm{yD}}=[\bar{\bar{Z}}_{\mathrm{eD}} \; \bar{\bar{Z}}_{\mathrm{hD}}]^{\mathrm{T}}$ of size $(N_{\mathrm{e}}+ N_{\mathrm{h}})\times N_{\mathrm{D}}$, and
\begin{equation}
\bar{\bar{Z}}_\mathrm{yy}=\begin{bmatrix}
\bar{\bar{Z}}_{\mathrm{ee}} & \bar{\bar{Z}}_{\mathrm{eh}} \\
\bar{\bar{Z}}_{\mathrm{he}} & \bar{\bar{Z}}_{\mathrm{hh}}
\end{bmatrix}
\end{equation}
of size $(N_{\mathrm{e}}+ N_{\mathrm{h}})\times (N_{\mathrm{e}}+N_{\mathrm{h}})$. Note that $\bar{\bar{Z}}_{\mathrm{Dy}}$ is dense while $\bar{\bar{Z}}_{\mathrm{yD}}$ and $\bar{\bar{Z}}_{\mathrm{yy}}$ are sparse. From the second row of Eq.~\eqref{eq:matrix_system2}, $\bar{I}_{\mathrm{y}}$ can be written in terms of $\bar{I}_{\mathrm{D}}$ as
\begin{equation}
\bar{I}_{\mathrm{y}} = -\bar{\bar{Z}}_{\mathrm{yy}}^{-1}\bar{\bar{Z}}_{\mathrm{yD}}\bar{I}_{\mathrm{D}}.
\label{eq:Iy}
\end{equation}
Substituting Eq.~\eqref{eq:Iy} into the first row of Eq.~\eqref{eq:matrix_system2} yields the reduced matrix system of size $N_{\mathrm{D}} \times N_{\mathrm{D}}$ as
\begin{equation}
(\bar{\bar{Z}}_{\mathrm{DD}}-\bar{\bar{Z}}_{\mathrm{Dy}} \bar{\bar{Z}}_{\mathrm{yy}}^{-1} \bar{\bar{Z}}_{\mathrm{yD}}) \bar{I}_{\mathrm{D}}=\bar{V}^{\mathrm{inc}}.
\label{eq:matrix_system_reduced}
\end{equation}
The reduced system is solved for $\bar{I}_{\mathrm{D}}$ using the transpose-free quasi-minimal residual (TFQMR) method~\cite{freund1993transpose}. Specifically, at each outer TFQMR iteration $(k)$, the matrix-vector product $(\bar{\bar{Z}}_{\mathrm{DD}} - \bar{\bar{Z}}_{\mathrm{Dy}}\bar{\bar{Z}}_{\mathrm{yy}}^{-1}\bar{\bar{Z}}_{\mathrm{yD}})\bar{I}^{(k)}_{\mathrm{D}}$ is never formed explicitly. Instead, $\bar{\bar{Z}}_{\mathrm{yy}}^{-1}\bar{\bar{Z}}_{\mathrm{yD}}\bar{I}^{(k)}_{\mathrm{D}}$ is computed by solving the sparse system $\bar{\bar{Z}}_{\mathrm{yy}}\tilde{\bar{I}}^{(k)}_\mathrm{y} = \bar{\bar{Z}}_{\mathrm{yD}}\bar{I}^{(k)}_{\mathrm{D}}$ using TFQMR at the inner level, after which the result is used to complete the matrix-vector product at the outer iteration~\cite{aibek2022solution}.

Compared with the conventional single-level iterative approach, which applies TFQMR directly to Eq.~\eqref{eq:matrix_system}, the two-level solver achieves significantly faster convergence as demonstrated in Section~\ref{sec:results}. This conclusion remains unchanged even when fast algorithms, such as the 
multilevel fast multipole algorithm~\cite{lu2003fast, nie2006fast, sertel2004mlfmm, takrimi2017incomplete} or the adaptive integral method~\cite{zhang2002vaim, nie2005precorrected, guo2006analysis}, are employed to accelerate matrix-vector products, since the main difference in computational cost arises from the number of iterations rather than the cost per iteration.

\section{Numerical Results}\label{sec:results}
In this section, several numerical examples are presented to demonstrate the accuracy, efficiency, and applicability of the proposed VIE-based solver for analyzing electromagnetic scattering from semiconductor nanostructures embedded in free space. In all examples, the excitation is a plane wave propagating along $\hat{\mathbf{z}}$ with an $\hat{\mathbf{x}}$-polarized electric field: 
\begin{align}
\mathbf{E}^{\mathrm{inc}}(\mathbf{r})=\hat{\mathbf{x}} E_{0} e^{-\mathrm{j} k_{0} z}.
\end{align}
Here, $E_0=1 \, \mathrm{V/m}$ is the amplitude of the incident electric field. All iterations of the two-level and single-level iterative solvers are terminated when the relative residual falls below the prescribed convergence threshold $\delta$: 
\begin{equation}
\label{eq:rel_res}
    \frac{\left\|\bar{b}-\bar{\bar{A}} \bar{I}^{(k)}\right\|_2}{\left\|\bar{b}\right\|_2}  \leq \delta
\end{equation}
where $\bar{b}$ is the right-hand side vector, $\bar{\bar{A}}$ is the system matrix, and $\bar{I}^{(k)}$ is the $k$th iterate. For the two-level iterative solver, the convergence thresholds for the inner and outer iterations are set to $\delta = 10^{-8}$ and $\delta = 10^{-4}$, respectively. For the single-level iterative solver, the convergence threshold is set to $\delta = 10^{-4}$.

All simulations are performed on Shaheen~II~\cite{shaheen2}, a Cray XC40 system at the KAUST Supercomputing Laboratory (KSL), consisting of compute nodes each equipped with two $16$-core Intel Haswell processors running at $2.3\,\mathrm{GHz}$ and $128\,\mathrm{GB}$ of DDR3 memory. The computationally intensive matrix operations and iterative solution procedure are parallelized using MPI and the simulations are executed on $128$ MPI processes distributed across $4$ compute nodes.
\subsection{Artificial Sphere}\label{sec:art_sphere}
In this example, electromagnetic scattering from a homogeneous nanosphere made of an artificial material is analyzed using the proposed VIE-based solver. The sphere has a radius of $10\,\mathrm{nm}$. The two-fluid HDE parameters of the artificial material are ${\omega_{\mathrm{e}}} = 3.6 \times {10^{14}}\,{\mathrm{rad/s}}$, $\omega_\mathrm{h} = 1.8 \times {10^{14}}\, {\mathrm{rad/s}}$, $\gamma_\mathrm{e} =\gamma_\mathrm{h} =1.0\times {10^{12}}\, {\mathrm{rad/s}}$, $\beta_{\mathrm{e}} = 4.3\times 10^5\, \mathrm{m/s}$, $\beta_{\mathrm{h}} =1.6\times 10^5 \, \mathrm{m/s}$, and $\varepsilon_{\mathrm{b}}(\mathbf{r})=5.0$~\cite{golestanizadeh2019hydrodynamic}.
% These semiconductor parameters are comparable to a realistic semiconductor parameters except the damping constant. In this artificial sphere example, the hole damping constant is set to a low value compared to realistic hole damping constant, since low damping constants make it easier to identify acoustic resonances at low frequencies. 
A total of $60$ simulations are performed using the two-level iterative solver at equally spaced frequency points over the range $\omega \in [0, 0.7\, \omega_{\mathrm{eff}}]$, where $\omega_{\mathrm{eff}}=({\omega_{\mathrm{e}}^2+\omega_{\mathrm{h}}^2})^{0.5}$. The nanostructure is discretized using a tetrahedral mesh, resulting in $N_\mathrm{D} = 17\,546$ and $N_\mathrm{e} = N_\mathrm{h}=16\,694$ basis functions. 

\begin{figure}[ht]
\centering
\subfigure[]{\includegraphics[width=0.45\columnwidth]{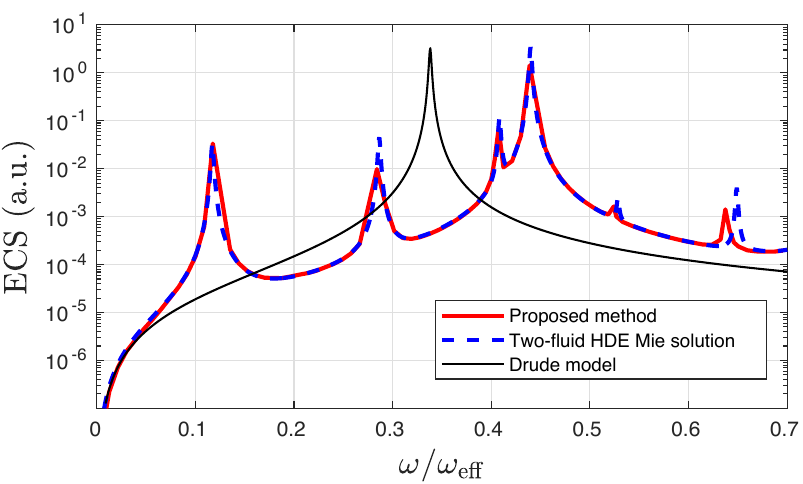}}\hspace{12pt}
\subfigure[]{\includegraphics[width=0.45\columnwidth]{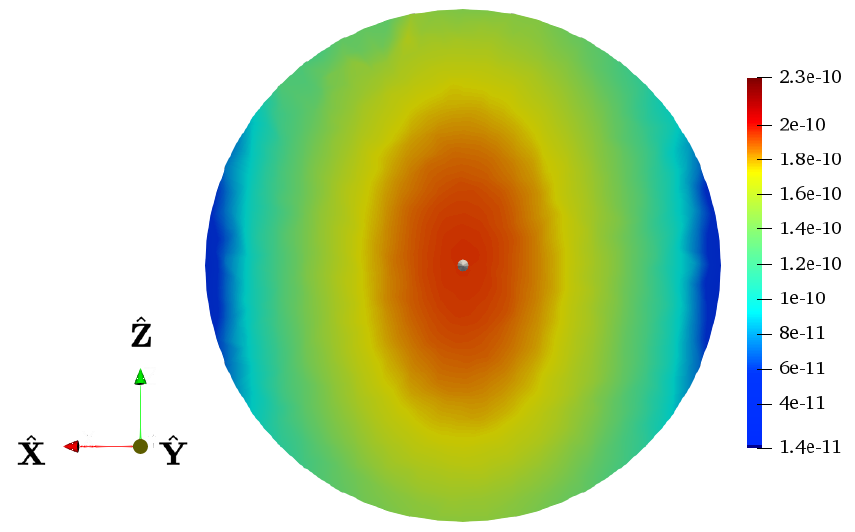}}\\
\subfigure[]{\includegraphics[width=0.45\columnwidth]{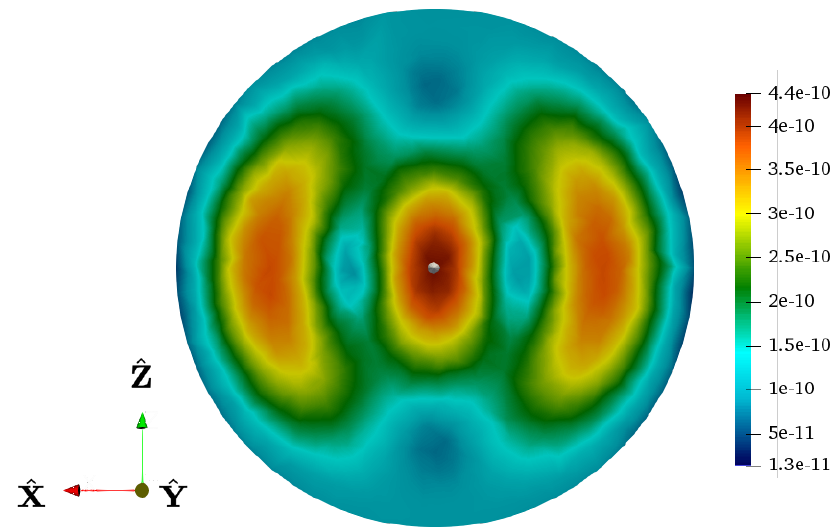}}\hspace{12pt}
\subfigure[]{\includegraphics[width=0.45\columnwidth]{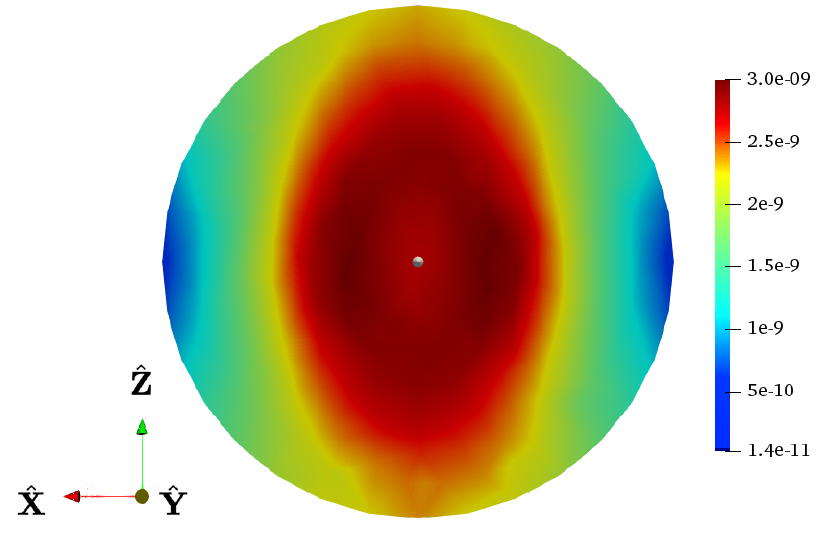}}
\caption{(a) ECS computed using the proposed VIE-based solver, the Mie series solution with the two-fluid HDE model, and the Mie series solution with the Drude model versus $\omega/\omega_\mathrm{eff}$. Field distributions on the $xz$-plane of the nanosphere at (b) $\omega = 0.11\,\omega_\mathrm{eff}$ (acoustic LSP resonance), (c) $\omega = 0.28\,\omega_\mathrm{eff}$ (first acoustic bulk resonance), and (d) $\omega = 0.53\,\omega_\mathrm{eff}$ (first optical bulk resonance associated with electrons).}
\label{fig:fig1}
\end{figure}

Fig.~\ref{fig:fig1}(a) presents the extinction cross section (ECS) as a function of $\omega/\omega_\mathrm{eff}$, computed using the proposed VIE-based solver, the Mie series solution with the two-fluid HDE model, and the Mie series solution with the Drude model~\cite{maack2018two}. Excellent agreement is observed between the proposed VIE-based solver and the Mie-series solution for the two-fluid HDE model, confirming the accuracy of the proposed VIE-based solver. The largest peak in all three ECS curves corresponds to the optical LSP resonance. This resonance appears at $\omega=0.33\, \omega_\mathrm{eff}$ in the Drude model result and shifts to a higher frequency $\omega=0.44\, \omega_\mathrm{eff}$ when the two-fluid HDE model is employed, indicating a blueshift of the resonance~\cite{aibek2022solution,zheng2018boundary}. Five additional peaks are observed at $\omega=0.11\, \omega_\mathrm{eff}$, $\omega=0.28\, \omega_\mathrm{eff}$, $\omega=0.41\, \omega_\mathrm{eff}$, $\omega=0.53\, \omega_\mathrm{eff}$, and $\omega=0.65\, \omega_\mathrm{eff}$ in the ECS computed using the proposed VIE-based solver and the Mie series solution with the two-fluid HDE model. Among these, the first peak corresponds to the acoustic LSP resonance, the second and third peaks correspond to acoustic bulk resonances, and the fourth and the fifth peaks correspond to optical bulk resonances associated with electrons and holes~\cite{golestanizadeh2019hydrodynamic}. In general, the optical bulk resonance induced by electrons is stronger than that induced by holes. To further illustrate these resonances, Figs.~\ref{fig:fig1}(b), (c), and (d) show the field distributions on the $xz$-plane of the nanosphere at $\omega = 0.11\,\omega_\mathrm{eff}$ (acoustic LSP resonance), $\omega = 0.28\,\omega_\mathrm{eff}$ (first acoustic bulk resonance), and $\omega = 0.53\,\omega_\mathrm{eff}$ (first optical bulk resonance associated with electrons), respectively.

\subsection{Semiconductor Sphere}\label{sec:InSb_sphere}
In this example, electromagnetic scattering from a nanosphere made of an indium antimonide ($\mathrm{InSb}$)-type semiconductor is analyzed using the proposed VIE-based solver. Four sets of simulations are performed to evaluate solver efficiency, validate the proposed solver against the two-fluid HDE Mie series solution, and investigate the effects of temperature and sphere radius on the scattering response. In all simulations, $\varepsilon_{\mathrm{b}}(\mathbf{r})=15.68$~\cite{dong2020modified}.

\begin{figure}[ht]
\centering
\subfigure[]{\includegraphics[width=0.48\columnwidth]{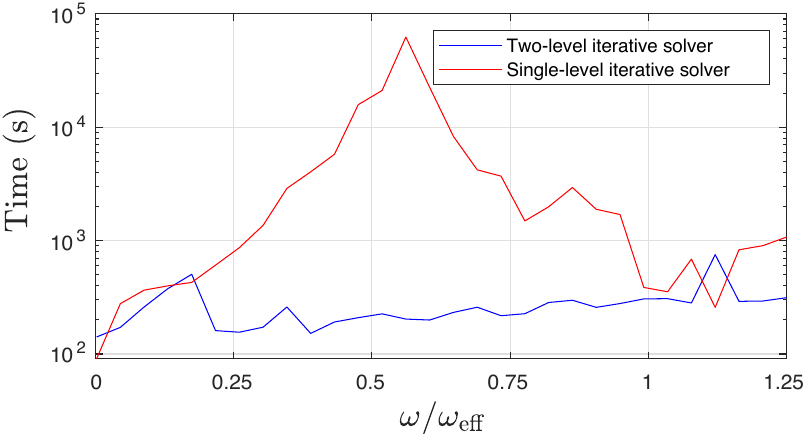}}\hspace{12pt}
\subfigure[]{\includegraphics[width=0.45\columnwidth]{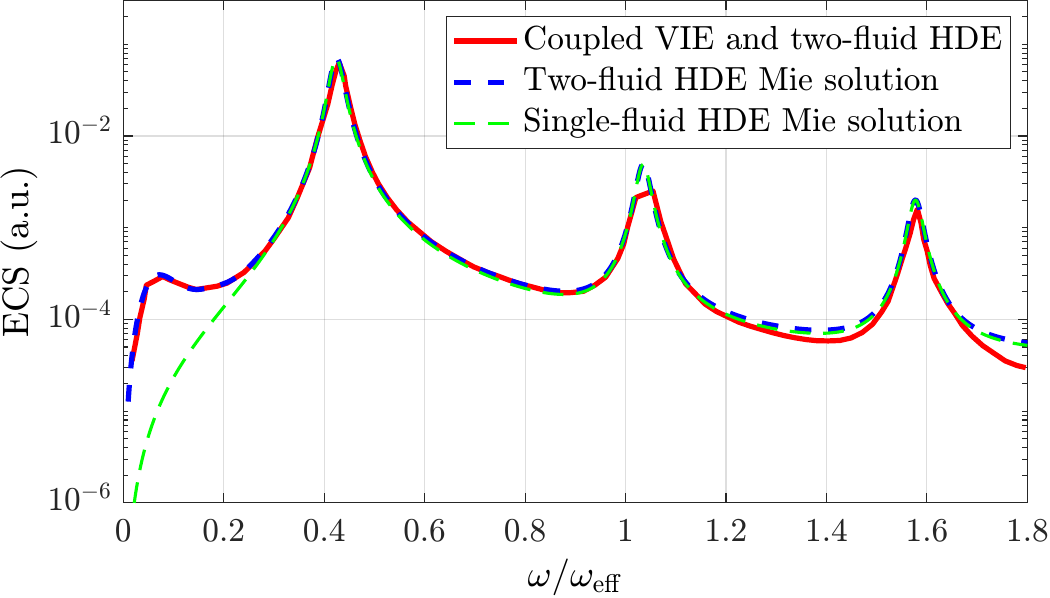}}\\
\subfigure[]{\includegraphics[width=0.48\columnwidth]{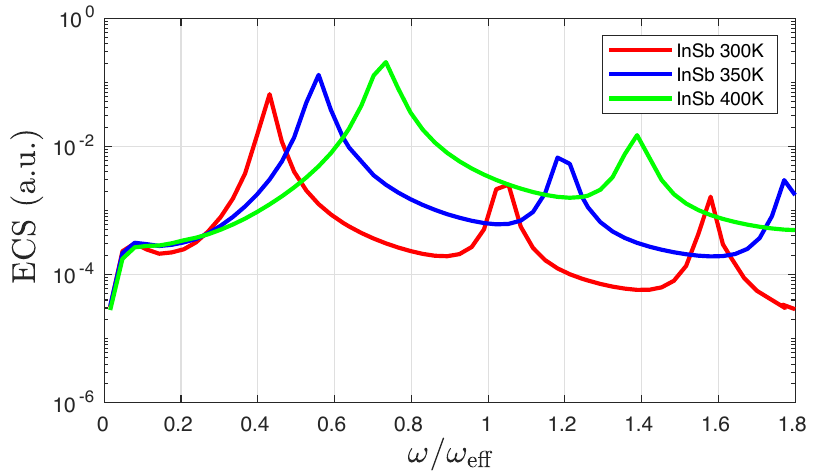}}\hspace{12pt}
\subfigure[]{\includegraphics[width=0.48\columnwidth]{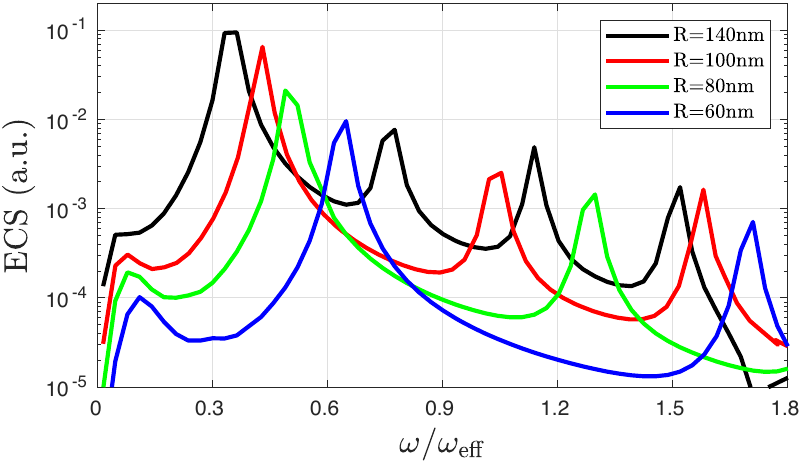}}
\caption{(a) Execution times of the single-level and two-level iterative solvers for the first set of simulations. (b) ECS computed using the proposed VIE-based solver, the Mie series solution with the two-fluid HDE model, and the Mie series solution with the single-fluid HDE model for the second set of simulations versus $\omega/\omega_\mathrm{eff}$. (c) ECS computed using the proposed VIE-based solver for three temperatures versus $\omega/\omega_\mathrm{eff}$. (d) ECS computed using the proposed VIE-based solver for four sphere radii versus $\omega/\omega_\mathrm{eff}$.}
\label{fig:fig2}
\end{figure}

In the first set of simulations, the sphere radius is $100\,\mathrm{nm}$ and the temperature is $300\, \mathrm{K}$. The two-fluid HDE parameters for $\mathrm{InSb}$ at $300\,\mathrm{K}$ are ${\omega_{\mathrm{e}}} = 6.34\times 10^{13}\,\mathrm{rad/s}$, $\omega_\mathrm{h} = 1.11\times 10^{13}\,\mathrm{rad/s}$, $\gamma_\mathrm{e} =1.99\times 10^{12}\,\mathrm{rad/s}$, $\gamma_\mathrm{h} =6.74\times 10^{12}\,\mathrm{rad/s}$, $\beta_{\mathrm{e}} = 10.89\times 10^{5}\, \mathrm{m/s}$, and $\beta_{\mathrm{h}} =1.92\times 10^{5} \, \mathrm{m/s}$~\cite{dong2020modified}. A total of $30$ simulations are carried out using both the single-level and two-level iterative solvers at equally spaced frequency points over the range $\omega \in [0, 1.25\,\omega_\mathrm{eff}]$. The nanostructure is discretized using a tetrahedral mesh with $N_\mathrm{D} = 9\,216$ and $N_\mathrm{e} = N_\mathrm{h}=8\,612$ basis functions. Fig.~\ref{fig:fig2}(a) compares the execution times of the single-level and two-level iterative solvers for this set of simulations. The results show that the two-level iterative solver is significantly faster on average. Both the single-level and two-level iterative solvers require approximately $3.9\,\mathrm{GB}$ of memory, confirming that the two-level solver achieves faster convergence without additional memory overhead.

In the second set of simulations, the sphere radius is $100\,\mathrm{nm}$ and the temperature is $300\, \mathrm{K}$. A total of $60$ simulations are carried out using the two-level iterative solver at equally spaced frequency points over the range $\omega \in [0, 1.8\, \omega_{\mathrm{eff}}]$. The nanostructure is discretized using a tetrahedral mesh with $N_\mathrm{D} = 17\,546$ and $N_\mathrm{e} = N_\mathrm{h}=16\,694$ basis functions. Fig.~\ref{fig:fig2}(b) shows the ECS computed using the proposed VIE-based solver, the Mie series solution with the two-fluid HDE model, and the Mie series solution with the single-fluid HDE model, confirming the accuracy of the proposed VIE-based solver for InSb-type semiconductors. Notably, the first resonance peak at $\omega = 0.07\,\omega_{\mathrm{eff}}$, corresponding to the acoustic LSP resonance~\cite{golestanizadeh2019hydrodynamic}, is not captured by the single-fluid HDE model.

In the third set of simulations, the sphere radius is $100\,\mathrm{nm}$ and three temperatures, $T \in \{ 300, 350, 400\}\,\mathrm{K}$, are considered. The two-fluid HDE parameters for $\mathrm{InSb}$ at each temperature are taken from~\cite{dong2020modified}. A total of $60$ simulations are carried out using the two-level iterative solver at equally spaced frequency points over the range $\omega \in [0, 1.8\, \omega_{\mathrm{eff}}]$. The nanostructure is discretized using a tetrahedral mesh with $N_\mathrm{D} = 17\,546$ and $N_\mathrm{e} = N_\mathrm{h}=16\,694$ basis functions. Fig.~\ref{fig:fig2}(c) shows the ECS computed using the proposed VIE-based solver for the three temperatures as a function of $\omega/\omega_\mathrm{eff}$. The resonance peak shifts toward higher frequencies and increases in amplitude as the temperature rises. This behavior can be explained by the increase in electron density in the conduction band and hole density in the valence band with increasing temperature.

In the fourth set of simulations, the sphere radius is varied over $R \in \{ 60, 80, 100, 140\}\,\mathrm{nm}$ and the temperature is $300\,\mathrm{K}$. The two-fluid HDE parameters are the same as those in the first set of simulations. For each radius, a total of $60$ simulations are carried out using the two-level iterative solver at equally spaced frequency points over the range $\omega \in [0, 1.8\,\omega_{\mathrm{eff}}]$, and the nanosphere is discretized using a tetrahedral mesh with $N_\mathrm{D}=17\,546$ and $N_\mathrm{e} = N_\mathrm{h}=16\,694$ basis functions. Fig.~\ref{fig:fig2}(d) shows the ECS computed using the proposed VIE-based solver for all four nanosphere radii as a function of $\omega/\omega_\mathrm{eff}$. The resonance peak shifts toward higher frequencies and decreases in amplitude as the sphere radius decreases.

% In addition, the LSP resonance peak at $\omega= XXX\omega_\mathrm{eff}$ and two other optical bulk resonance peaks at $\omega= 1.03\omega_\mathrm{eff}$ and $\omega= 1.15\omega_\mathrm{eff}$ are captured by all three methods. 

% In this figure, we only see acoustic LPS and but not acoustic bulk peaks, since hole damping constant (calculated using hole mobility) is very high as mentioned above.  

% Acturally, similar blue shift phenomenon also happens for the single-fluid HDE model. 

\subsection{Semiconductor Dimer}\label{sec:dimer}
\begin{figure}[ht]
\centering
\subfigure[]{\includegraphics[width=0.55\columnwidth]{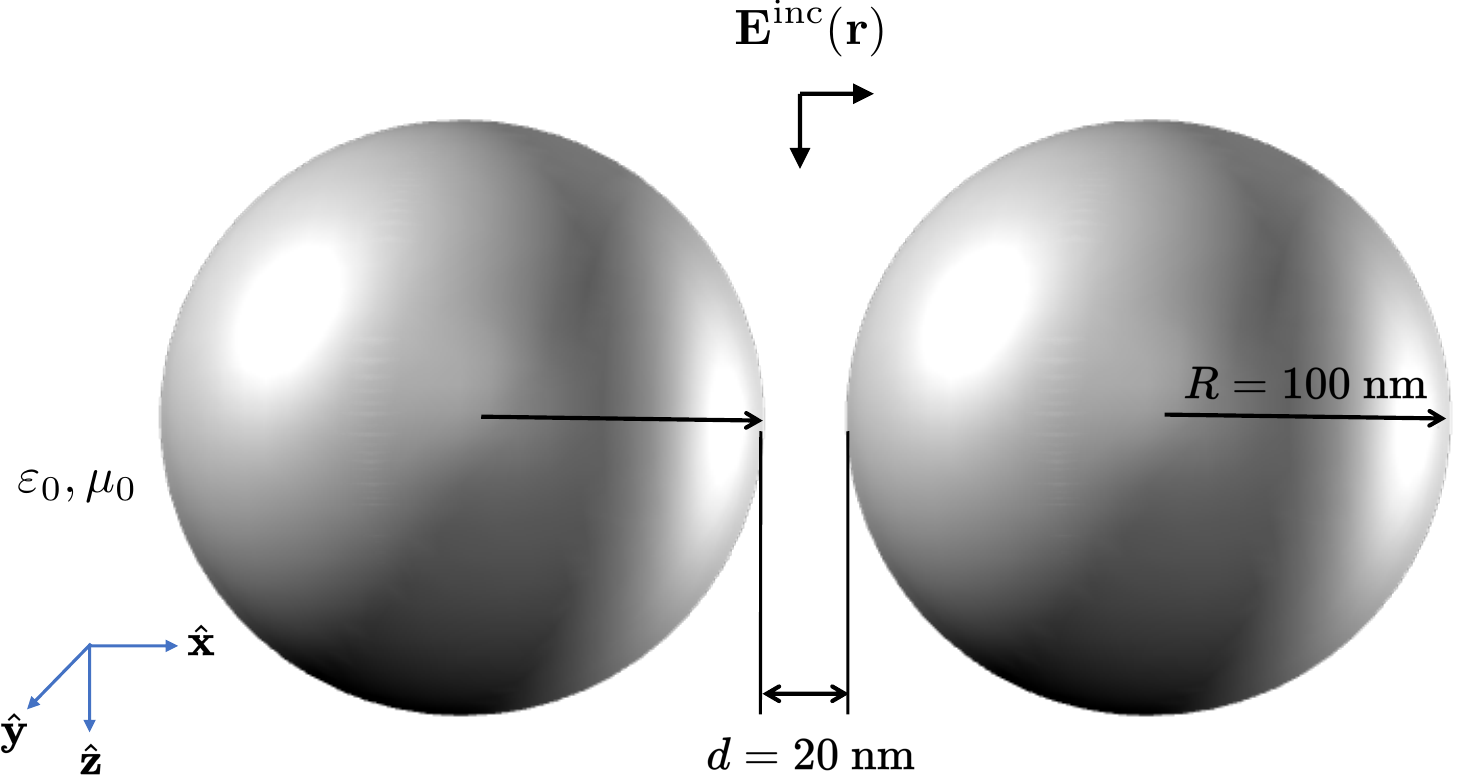}}\hspace{12pt}\\
\subfigure[]{\includegraphics[width=0.58\columnwidth]{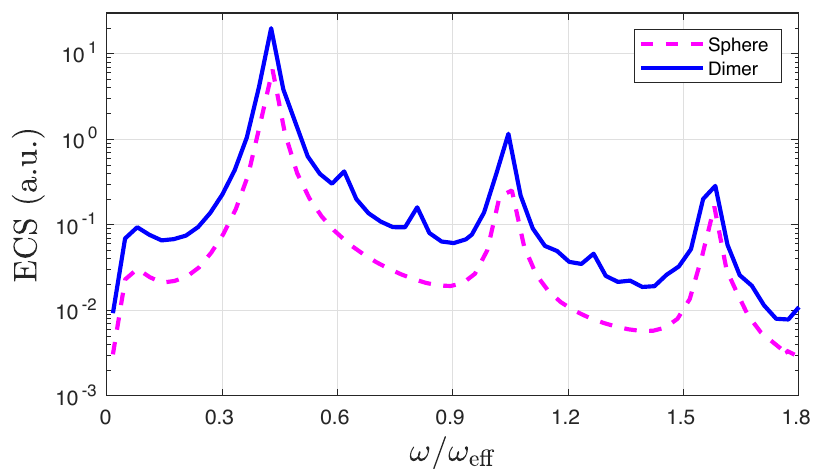}}
\caption{(a) Geometry of the scattering problem involving the semiconductor nanodimer. (b) ECS computed using the two-level iterative solver for two scattering scenarios (the nanodimer and the single nanosphere) versus $\omega/\omega_\mathrm{eff}$.}
\label{fig:fig3}
\end{figure}

In this example, electromagnetic scattering from a nanodimer made of an $\mathrm{InSb}$-type semiconductor [see Fig.~\ref{fig:fig3}(a)] is analyzed using the proposed VIE-based solver. Each sphere has a radius of $100\, \mathrm{nm}$ and the shortest distance between the spheres is $20\,\mathrm{nm}$. The temperature is $300\,\mathrm{K}$ and the two-fluid HDE parameters are those of $\mathrm{InSb}$ at $300\,\mathrm{K}$ used in the first set of simulations in Section~\ref{sec:InSb_sphere}. Two scattering scenarios are considered: (i) the nanodimer, in which both spheres are discretized using the same tetrahedral mesh, resulting in $N_\mathrm{D} = 18\,186$ and $N_\mathrm{e} = N_\mathrm{h} = 16\,978$ basis functions for the full structure, and (ii) a single nanosphere of radius $100\,\mathrm{nm}$, discretized with $N_\mathrm{D} = 17\,546$ and $N_\mathrm{e} = N_\mathrm{h} = 16\,694$ basis functions. For both scenarios, $60$ simulations are carried out using the two-level iterative solver at equally spaced frequency points over the range $\omega \in [0, 1.8\,\omega_{\mathrm{eff}}]$.

Fig.~\ref{fig:fig3}(b) shows the ECS computed for the two scattering scenarios described above as a function of $\omega/\omega_\mathrm{eff}$. The ECS of the nanodimer is larger than that of the single nanosphere over the entire frequency range due to the electromagnetic coupling between the two spheres, which leads to stronger field enhancement. As reported 
in~\cite{golestanizadeh2019hydrodynamic}, the ECS of the nanodimer increases as the separation between the two spheres decreases. The main resonance peaks at $\omega=0.07\, \omega_\mathrm{eff}$, $\omega=0.44\, \omega_\mathrm{eff}$, $\omega=1.04\, \omega_\mathrm{eff}$, and $\omega=1.59\, \omega_\mathrm{eff}$ are observed in both ECS curves. In addition, several minor resonance peaks are present in the ECS of the nanodimer, which are attributed to higher-order modes excited by the near-field coupling between the two spheres~\cite{maack2018two}.

\subsection{Semiconductor Cylinder}\label{sec:cylinder}
\begin{figure}[ht]
\centering
\subfigure[]{\includegraphics[width=0.48\columnwidth]{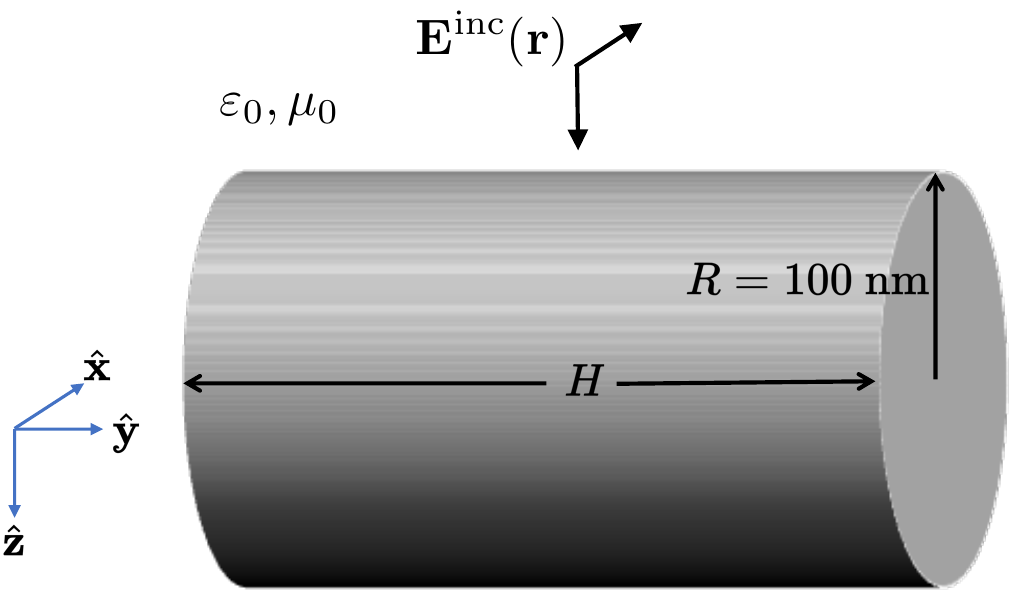}}\hspace{12pt}
\subfigure[]{\includegraphics[width=0.48\columnwidth]{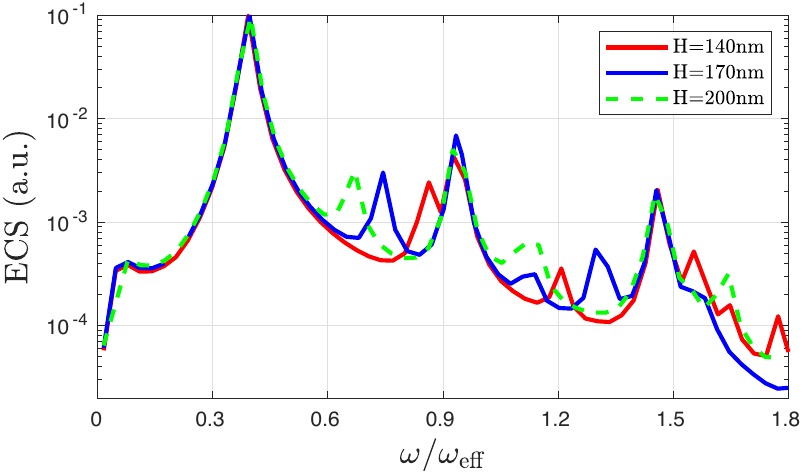}}\\
\subfigure[]{\includegraphics[width=0.3\columnwidth]{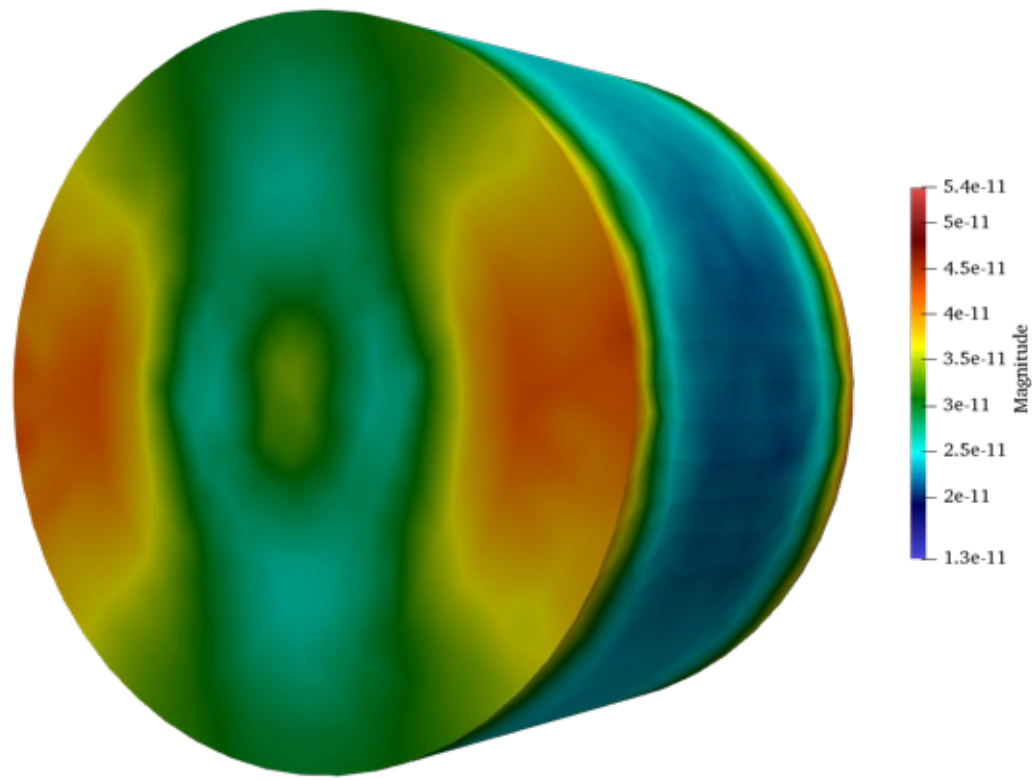}}\hspace{12pt}
\subfigure[]{\includegraphics[width=0.3\columnwidth]{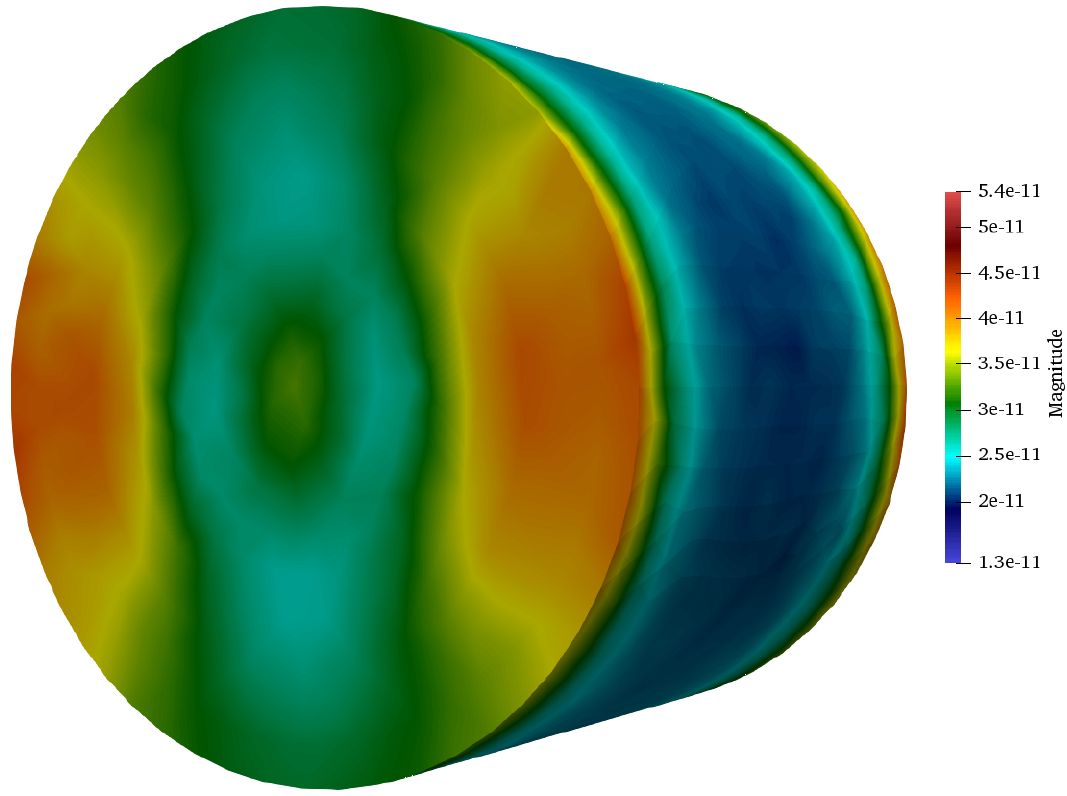}}\hspace{12pt}
\subfigure[]{\includegraphics[width=0.3\columnwidth]{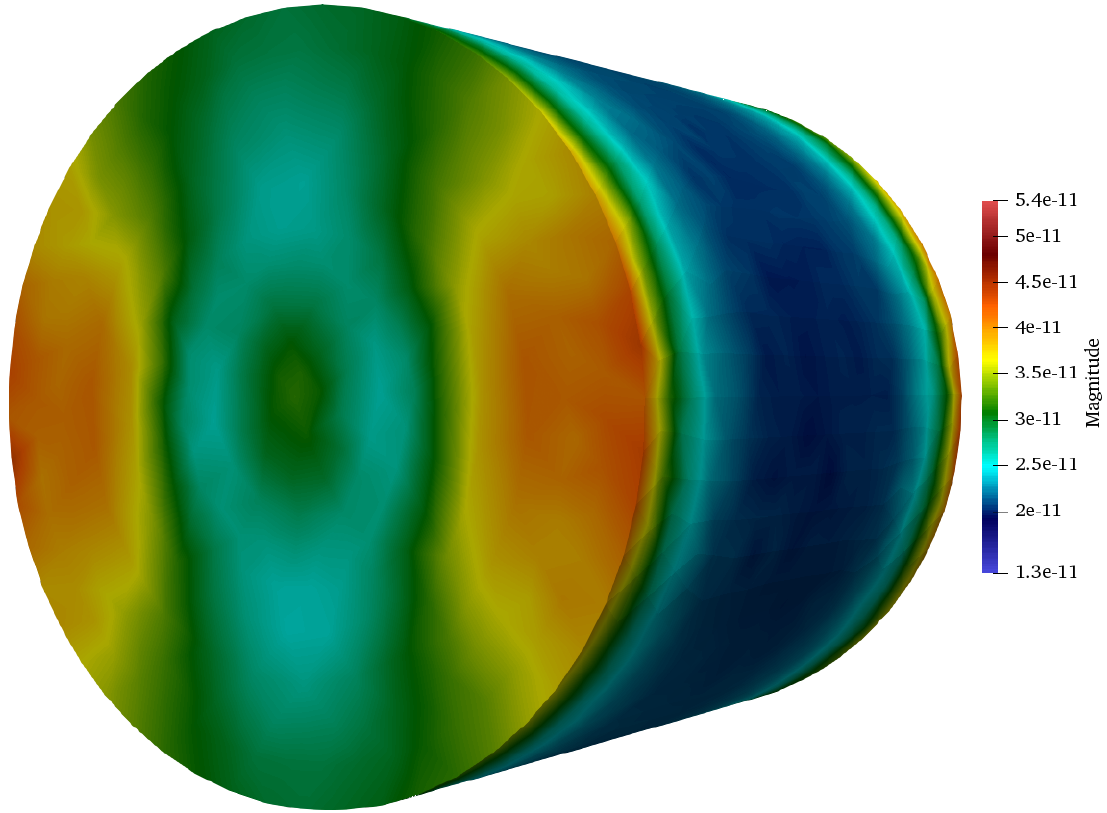}}\\
\subfigure[]{\includegraphics[width=0.3\columnwidth]{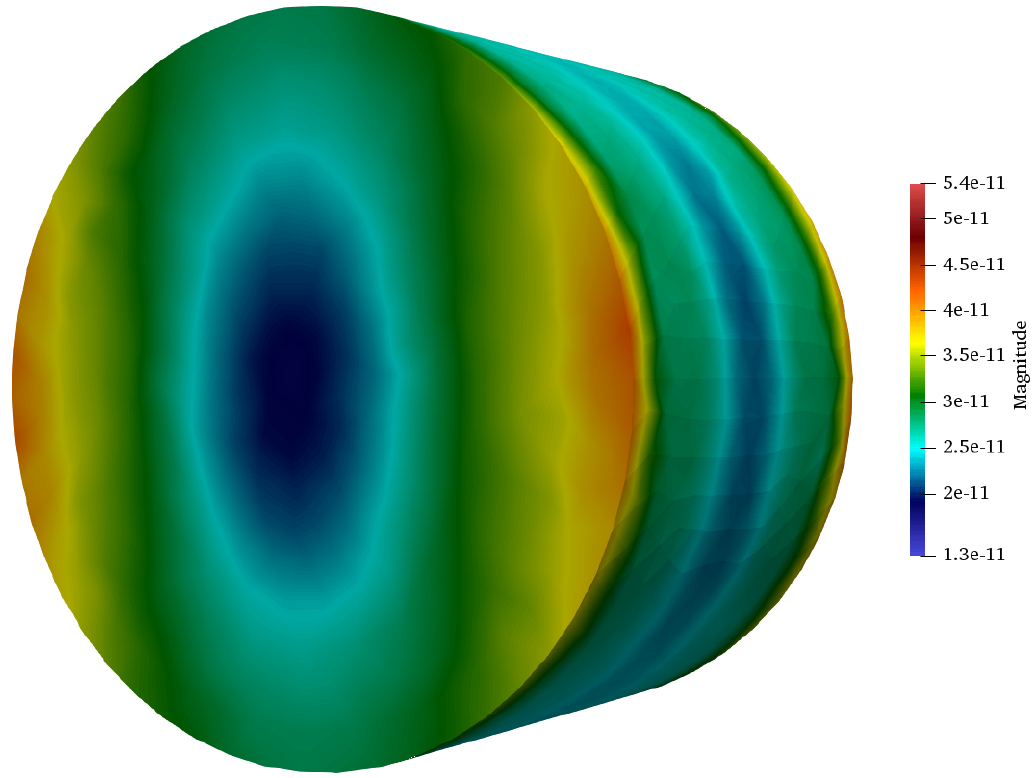}}\hspace{12pt}
\subfigure[]{\includegraphics[width=0.3\columnwidth]{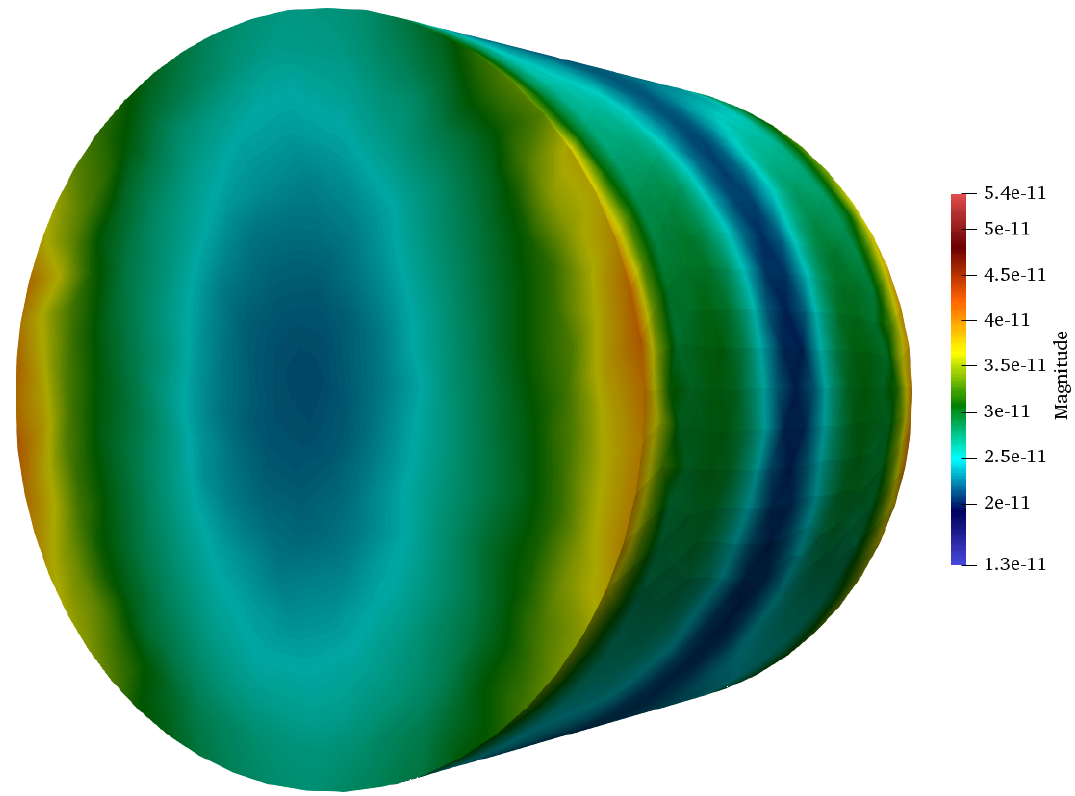}}\hspace{12pt}
\subfigure[]{\includegraphics[width=0.3\columnwidth]{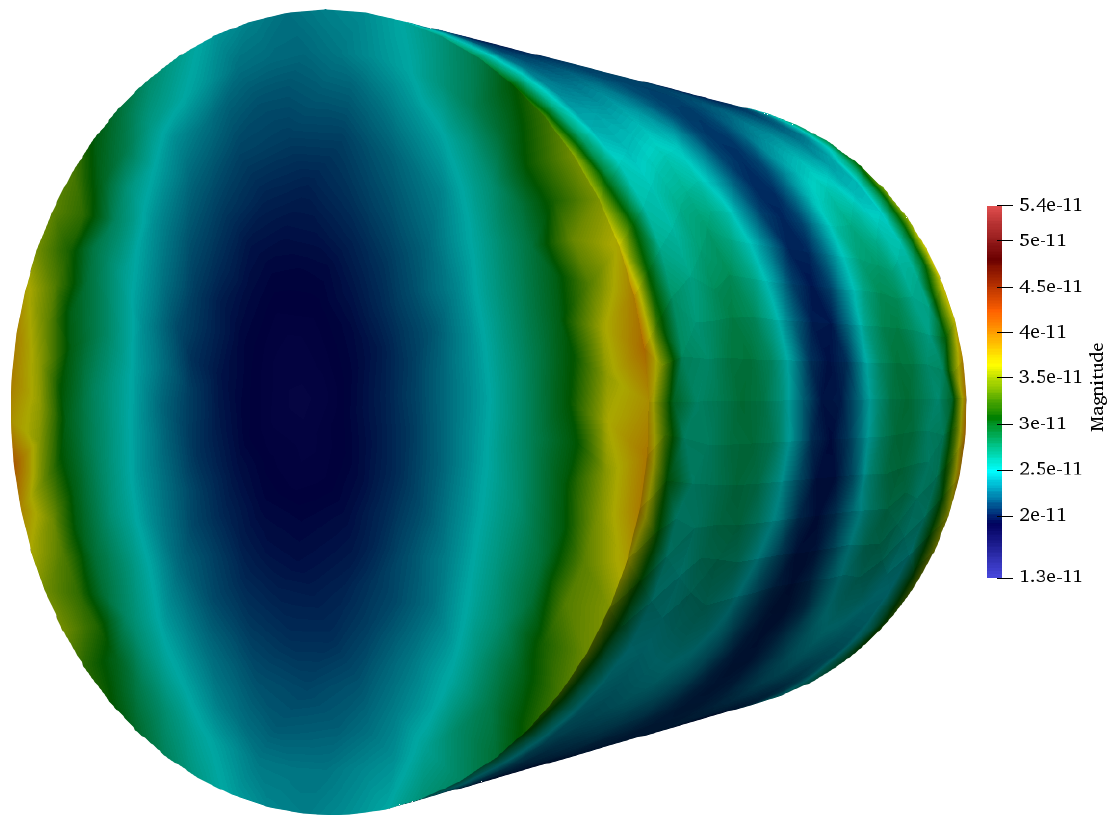}}
\caption{(a) Geometry of the scattering problem involving the semiconductor nanocylinder. (b) ECS computed using the two-level 
iterative solver for three nanocylinder heights versus $\omega/\omega_\mathrm{eff}$. Field distributions on the $xz$-plane of the nanocylinder with height (c) $H=140\,\mathrm{nm}$, (d) $H=170\,\mathrm{nm}$, and (e) $H=200\,\mathrm{nm}$ for the first optical bulk resonance. Field distributions on the $xz$-plane of the nanocylinder with height (f) $H=140\,\mathrm{nm}$, (g) $H=170\,\mathrm{nm}$, and (h) $H=200\,\mathrm{nm}$ for the first optical LSP resonance associated with height.}
\label{fig:fig4}
\end{figure}

In this example, electromagnetic scattering from a nanocylinder made of an $\mathrm{InSb}$-type semiconductor is analyzed using the proposed VIE-based solver for different cylinder heights [see Fig.~\ref{fig:fig4}(a)]. The cylinder height is varied over $H \in \{140, 170, 200\}\,\mathrm{nm}$ and the radius is $100\,\mathrm{nm}$. The temperature is $300\,\mathrm{K}$ and the two-fluid HDE parameters are those of $\mathrm{InSb}$ at $300\,\mathrm{K}$ used in the first set of simulations in Section~\ref{sec:InSb_sphere}. For each nanocylinder, $60$ simulations are carried out using the two-level iterative solver at equally spaced frequency points over the range $\omega \in [0, 1.8\, \omega_{\mathrm{eff}}]$. Each nanocylinder is discretized using a tetrahedral mesh, resulting in $N_{\mathrm{D}} \in \{18\,476, 22\,482, 26\,138\}$ and $N_{\mathrm{e}} = N_{\mathrm{h}} \in \{17\,476, 21\,354, 24\,914\}$ basis functions for $H \in \{140, 170, 200\}\,\mathrm{nm}$, respectively.

Fig.~\ref{fig:fig4}(b) shows the ECS computed using the two-level iterative solver for three nanocylinder heights as a function of $\omega/\omega_{\mathrm{eff}}$. The main resonance peaks at $\omega = 0.41\,\omega_\mathrm{eff}$, $\omega =0.93\,\omega_\mathrm{eff}$, and $\omega = 1.47\,\omega_\mathrm{eff}$ are observed in all three ECS curves, while the remaining resonance peaks shift toward higher frequencies as the cylinder height decreases. Notably, these height-dependent resonances are excited along the $\hat{\mathbf{y}}$-direction even though $\mathbf{E}^{\mathrm{inc}}(\mathbf{r})$ is polarized along $\hat{\mathbf{x}}$. By analogy with the sphere results in Section~\ref{sec:art_sphere}, the peak at $\omega = 0.41\,\omega_{\mathrm{eff}}$ corresponds to the first optical bulk resonance. Figs.~\ref{fig:fig4}(c)--(e) show the field distributions at this peak for all three nanocylinder heights. Similarly, the first height-dependent peak corresponds to the first optical LSP resonance, and Figs.~\ref{fig:fig4}(f)--(h) show the corresponding field distributions.

% In Fig.~\ref{fig:fig4} (b), one can see that all the main peaks (e.g., optical LSP, longitudinal resonances, acoustic LSP) are almost appeared at  same frequencies. 

% This might be caused due to the divergence operator in (eqXXX). 

% There are many other peaks corresponding to the second and third resonance points of the cylinders. However, it is hard to distinguish them, since they are coupled with the main optical bulk resonance peaks

\subsection{Semiconductor Hexagonal Prism}\label{sec:prism}
\begin{figure}[ht]
\centering
\subfigure[]{\includegraphics[width=0.58\columnwidth]{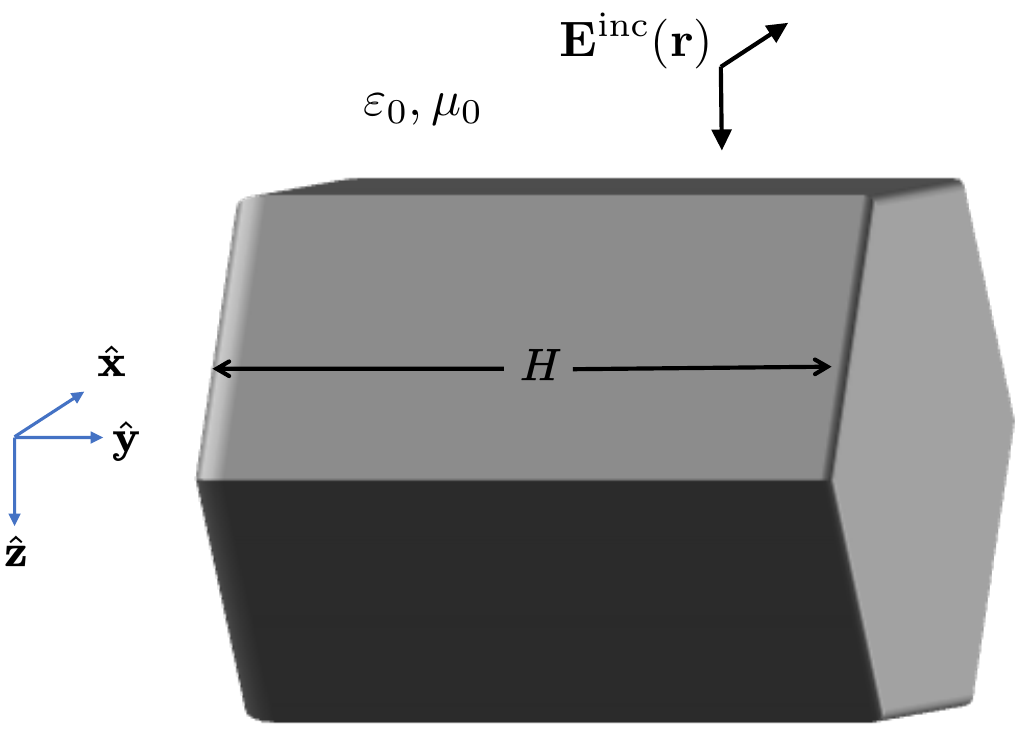}}\hspace{12pt}\\
\subfigure[]{\includegraphics[width=0.58\columnwidth]{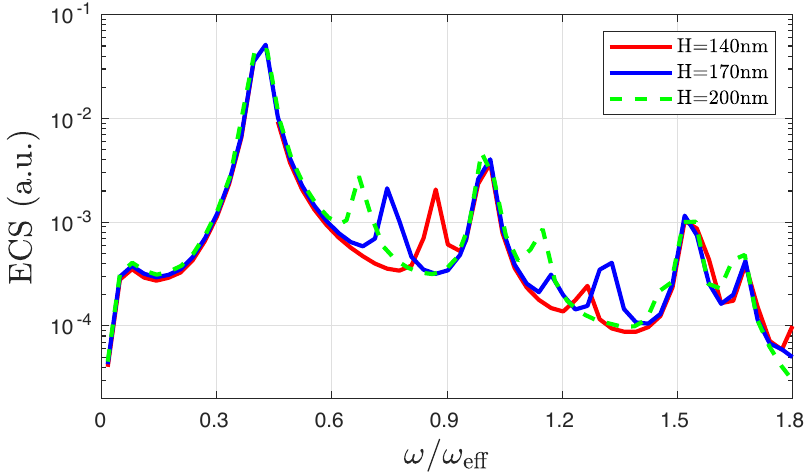}}
\caption{(a) Geometry of the scattering problem involving the semiconductor nanoprism. (b) ECS computed using the two-level iterative solver for the three prism heights versus $\omega/\omega_\mathrm{eff}$.}
\label{fig:fig5}
\end{figure}

In this example, electromagnetic scattering from an $\mathrm{InSb}$-type semiconductor hexagonal nanoprism is analyzed using the proposed VIE-based solver for different prism heights [see Fig.~\ref{fig:fig5}(a)]. The prism height is varied over $H \in \{140, 170, 200\} \,\mathrm{nm}$ and the edge length of the hexagonal cross section is $100\,\mathrm{nm}$. The temperature is $300\,\mathrm{K}$ and the two-fluid HDE parameters are those of $\mathrm{InSb}$ at $300\,\mathrm{K}$ used in the first set of simulations in Section~\ref{sec:InSb_sphere}. For each nanoprism, $60$ simulations are carried out using the two-level iterative solver at equally spaced frequency points over the range $\omega \in [0, 1.8\,\omega_{\mathrm{eff}}]$. Each nanoprism is discretized using a tetrahedral mesh, resulting in $N_{\mathrm{D}} \in \{14\,846, 18\,132, 21\,422\}$ and $N_{\mathrm{e}} = N_{\mathrm{h}} \in \{13\,994, 17\,136, 20\,354\}$ basis functions for $H \in \{140, 170, 200\}\,\mathrm{nm}$, respectively.

Fig.~\ref{fig:fig5}(b) shows the ECS computed using the two-level iterative solver for the three nanoprism heights as a function of 
$\omega/\omega_{\mathrm{eff}}$. By analogy with the sphere results in Section~\ref{sec:art_sphere}, the peak at $\omega = 0.41\,\omega_{\mathrm{eff}}$, observed in all three ECS curves, corresponds to the first optical bulk resonance, while the remaining peaks shift toward higher frequencies as the prism height decreases, consistent with the height-dependent optical LSP resonances observed in Section~\ref{sec:cylinder}. 

% Note that this hexagonal prism is often fabricated for crystal structures~\cite{lu2006semiconductor}. 

% Fig.~\ref{fig:fig5}(c) compares ECS for the cylinder to that for the hexagonal prism with height $140\,nm$. As seen from this figure, ECS for the hexagonal prism is blue shifted comparing to the cylinder. This is due to the fact that the size of the hexagonal prisms is XXX. Although the heights are same, the area of the circle and the hexagon are different and the difference in the area of the circle for a given example is higher. It also means that two-fluid HDE models are sensitive to the size dependent models and even small changes in the are might affect the ECS [ref].

\section{Conclusion}

Electromagnetic scattering from semiconductor nanostructures is analyzed using a coupled system of the VIE and the two-fluid HDE. The VIE expresses the electric field as a convolution of the scalar Green function of the background medium with the electric flux density and free-electron and hole polarization currents, while the two-fluid HDE relates the electric field to these polarization currents and accounts for nonlocal dispersion effects.

For the numerical solution of this coupled system, the nanostructure is discretized using a tetrahedral mesh, and the unknown electric flux density and free-electron and hole polarization currents are expanded using SWG basis functions defined on the mesh. Substituting these expansions into the coupled VIE and two-fluid HDE and applying Galerkin testing yield a matrix system for the unknown expansion coefficients, which is solved efficiently using a two-level iterative solver. Numerical results for InSb-type semiconductor nanostructures demonstrate the accuracy and efficiency of the proposed VIE-based solver and its ability to capture unique optical phenomena, such as acoustic plasmon resonances and the blueshift of LSP resonances, that cannot be described by the single-fluid HDE or classical Drude-based models. Future work will focus on extending the solver to excitation-free simulations for characteristic mode analysis of semiconductor nanostructures.  

\section{Appendix}
The matrix entries of the coupled system in Eq.~\eqref{eq:matrix_system} are derived following the procedure presented in~\cite{aibek2022solution} for the coupled system of the VIE and the single-fluid HDE. Since the free-electron and hole polarization currents enter the coupled system symmetrically, the blocks $\bar{\bar{Z}}_{\mathrm{Dq}}$, $\bar{\bar{Z}}_{\mathrm{qD}}$, $\bar{\bar{Z}}_{\mathrm{qq}}$, and $\bar{\bar{Z}}_{\mathrm{qp}}$ can be derived simultaneously using unified indices $\mathrm{q}, \mathrm{p} \in \{\mathrm{e}, \mathrm{h}\}$, $\mathrm{q}\neq\mathrm{p}$.

The entries of all blocks and the excitation vector $\bar{V}^{\mathrm{inc}}$ are given by
\begin{equation}
\{\bar{\bar{Z}}_{\mathrm{DD}}\}_{mn}=\frac{1}{\varepsilon_{0}}
\Big\langle\mathbf{f}_{m}^{\mathrm{D}}(\mathbf{r}), 
\frac{\mathbf{f}_{n}^{\mathrm{D}}(\mathbf{r})}{\varepsilon_{\mathrm{b}}
(\mathbf{r})}\Big\rangle -\mathrm{j} \omega\Big\langle\mathbf{f}_{m}^{\mathrm{D}}
(\mathbf{r}), \mathcal{L}[\kappa \mathbf{f}_{n}^{\mathrm{D}}]
(\mathbf{r})\Big\rangle 
\label{eq:ZDD}
\end{equation}
\begin{equation}
\{\bar{\bar{Z}}_{\mathrm{Dq}}\}_{mn}=-\frac{1}{\mathrm{j} \omega 
\varepsilon_{0}}\Big\langle\mathbf{f}_{m}^{\mathrm{D}}(\mathbf{r}), 
\frac{\mathbf{f}_{n}^{\mathrm{q}}(\mathbf{r})}{\varepsilon_{\mathrm{b}}
(\mathbf{r})}\Big\rangle-\Big\langle\mathbf{f}_{m}^{\mathrm{D}}
(\mathbf{r}), \mathcal{L}[\frac{\mathbf{f}_{n}^{\mathrm{q}}}
{\varepsilon_{\mathrm{b}}}](\mathbf{r})\Big\rangle 
\label{eq:ZDq}
\end{equation}
\begin{equation}
\{\bar{\bar{Z}}_{\mathrm{qD}}\}_{mn}=\mathrm{j} \omega \omega_{\mathrm{q}}^{2}
\Big\langle\mathbf{f}_{m}^{\mathrm{q}}(\mathbf{r}), 
\frac{\mathbf{f}_{n}^{\mathrm{D}}(\mathbf{r})}{\varepsilon_{\mathrm{b}}
(\mathbf{r})}\Big\rangle 
\label{eq:ZqD}
\end{equation}
\begin{equation}
\{\bar{\bar{Z}}_{\mathrm{qq}}\}_{mn}=\beta^{2}_\mathrm{q}
\Big\langle\mathbf{f}_{m}^{\mathrm{q}}(\mathbf{r}), \nabla[\nabla 
\cdot \mathbf{f}_{n}^{\mathrm{q}}(\mathbf{r})]\Big\rangle 
+\omega(\omega-\mathrm{j} \gamma_\mathrm{q})\Big\langle\mathbf{f}_{m}^{\mathrm{q}}
(\mathbf{r}), \mathbf{f}_{n}^{\mathrm{q}}(\mathbf{r})\Big\rangle
-\omega_{\mathrm{q}}^{2}\Big\langle\mathbf{f}_{m}^{\mathrm{q}}(\mathbf{r}), 
\frac{\mathbf{f}_{n}^{\mathrm{q}}(\mathbf{r})}{\varepsilon_{\mathrm{b}}
(\mathbf{r})}\Big\rangle 
\label{eq:Zqq}
\end{equation}
\begin{equation}
\{\bar{\bar{Z}}_{\mathrm{qp}}\}_{mn}=-\omega_{\mathrm{q}}^{2}
\Big\langle\mathbf{f}_{m}^{\mathrm{q}}(\mathbf{r}), 
\frac{\mathbf{f}_{n}^{\mathrm{p}}(\mathbf{r})}{\varepsilon_{\mathrm{b}}
(\mathbf{r})}\Big\rangle 
\label{eq:Zqp}
\end{equation}
\begin{equation}
\{\bar{V}^{\mathrm{inc}}\}_{m}=\Big\langle
\mathbf{f}_{m}^{\mathrm{D}}(\mathbf{r}), 
\mathbf{E}^{\mathrm{inc}}(\mathbf{r})\Big\rangle.
\label{eq:Vinc}
\end{equation}
Here, the inner product is defined as
\begin{equation*}
\Big\langle \mathbf{a}(\mathbf{r}), \mathbf{b}(\mathbf{r}) \Big\rangle = 
\int_{V_{\mathrm{a}} \cap \, V_{\mathrm{b}}} \mathbf{a}(\mathbf{r}) 
\cdot \mathbf{b}(\mathbf{r}) \, dv
\end{equation*} 
where $V_{\mathrm{a}}$ and $V_{\mathrm{b}}$ are the supports of $\mathbf{a}(\mathbf{r})$ and $\mathbf{b}(\mathbf{r})$, respectively.

As seen from Eqs.~\eqref{eq:ZDD} and~\eqref{eq:ZDq}, only $\bar{\bar{Z}}_{\mathrm{DD}}$ and $\bar{\bar{Z}}_{\mathrm{Dq}}$ involve the operator $\mathcal{L}[\cdot](\cdot)$ and are therefore dense. The remaining blocks $\bar{\bar{Z}}_{\mathrm{qD}}$, $\bar{\bar{Z}}_{\mathrm{qq}}$, and $\bar{\bar{Z}}_{\mathrm{qp}}$ are sparse, each having at most seven nonzero entries per row, as can be verified from the explicit expressions provided in Sections~\ref{sec:ZqD}--\ref{sec:Zqp}. In the derivation of the explicit expressions, $\varepsilon_{\mathrm{b}}(\mathbf{r})$ and $\kappa(\mathbf{r})$ are assumed to be constant within each tetrahedron, with values obtained by sampling at the center of that tetrahedron. Accordingly, local functions are defined as
\begin{equation}
\varepsilon_{\mathrm{b},n}(\mathbf{r}) = \begin{cases} 
\varepsilon_{\mathrm{b},n}^{+} = \varepsilon_{\mathrm{b}}
(\mathbf{r}_{\mathrm{c}}^{+}), & \mathbf{r} \in V_n^{+} \\ 
\varepsilon_{\mathrm{b},n}^{-} = \varepsilon_{\mathrm{b}}
(\mathbf{r}_{\mathrm{c}}^{-}), & \mathbf{r} \in V_n^{-} 
\end{cases}
\label{eq:eps_local}
\end{equation}
\begin{equation}
\kappa_n(\mathbf{r}) = \begin{cases} 
\kappa_n^{+} = \kappa(\mathbf{r}_{\mathrm{c}}^{+}), 
& \mathbf{r} \in V_n^{+} \\ 
\kappa_n^{-} = \kappa(\mathbf{r}_{\mathrm{c}}^{-}), 
& \mathbf{r} \in V_n^{-} 
\end{cases}
\label{eq:kappa_local}
\end{equation}
where $\mathbf{r}_{\mathrm{c}}^{\pm}$ are the centers of $V_n^{\pm}$. In the following sections, the explicit expressions for the entries of all blocks are provided for all combinations of full and half SWG functions used as testing and basis functions.

\subsection{Entries of $\bar{\bar{Z}}_{\mathrm{DD}}$}
\label{sec:ZDD}
Four possible combinations of the testing function $\mathbf{f}_m^{\mathrm{D}}(\mathbf{r})$ and the basis function $\mathbf{f}_n^{\mathrm{D}}(\mathbf{r})$ must be considered, each leading to a different expression for the matrix entries.

\textbf{Case 1:} Both $\mathbf{f}_m^{\mathrm{D}}(\mathbf{r})$ and $\mathbf{f}_n^{\mathrm{D}}(\mathbf{r})$ are full SWG functions:
\begin{equation}
\label{eq:ZDD_ff}
\begin{aligned}
 \{\bar{\bar{Z}}_{\mathrm{DD}}\}_{m n}=&\frac{1}{\varepsilon_0} \int_{V_m} \frac{\mathbf{f}_m^{\mathrm{D}}(\mathbf{r}) \cdot \mathbf{f}_n^{\mathrm{D}}(\mathbf{r})}{\varepsilon_{\mathrm{b}, n}(\mathbf{r})}\,dv \\
& -\omega^2 \mu_0 \int_{V_m} \mathbf{f}_m^{\mathrm{D}}(\mathbf{r}) \cdot \int_{V_n} \kappa_n(\mathbf{r}^{\prime}) \mathbf{f}_n^{\mathrm{D}}(\mathbf{r}^{\prime}) G(\mathbf{r}, \mathbf{r}^{\prime})\,dv^{\prime}\,dv \\
& +\frac{1}{\varepsilon_0} \int_{V_m} \nabla \cdot \mathbf{f}_m^{\mathrm{D}}(\mathbf{r})\left\{\int_{V_n} \kappa_n(\mathbf{r}^{\prime}) \nabla^{\prime} \cdot \mathbf{f}_n^{\mathrm{D}}(\mathbf{r}^{\prime}) G(\mathbf{r}, \mathbf{r}^{\prime})\,dv^{\prime}\right. \\
&\qquad\qquad\qquad\qquad\qquad\qquad \left. -(\kappa_n^{+}-\kappa_n^{-}) \int_{S_n} G(\mathbf{r}, \mathbf{r}^{\prime})\,d s^{\prime}\right\}\,dv.
\end{aligned}
\end{equation}

\textbf{Case 2:} $\mathbf{f}_m^{\mathrm{D}}(\mathbf{r})$ is a half SWG function and $\mathbf{f}_n^\mathrm{D}(\mathbf{r})$ is a full SWG function:
\begin{equation}
\label{eq:ZDD_hf}
\begin{aligned}
\{\bar{\bar{Z}}_\mathrm{DD}\}_{mn} &=\frac{1}{\varepsilon_0} \int_{V_m^{+}} \frac{\mathbf{f}_m^{\mathrm{D}}(\mathbf{r}) \cdot 
\mathbf{f}_n^{\mathrm{D}}(\mathbf{r})}{\varepsilon_{\mathrm{b}, n}(\mathbf{r})}\,dv \\
& -\omega^2 \mu_0 \int_{V_m^{+}} \mathbf{f}_m^{\mathrm{D}}(\mathbf{r}) \cdot \int_{V_n} \kappa_n(\mathbf{r}^{\prime}) 
\mathbf{f}_n^{\mathrm{D}}(\mathbf{r}^{\prime}) G(\mathbf{r}, \mathbf{r}^{\prime})\,dv^{\prime}\,dv \\
& +\frac{1}{\varepsilon_0} \int_{V_m^{+}} \nabla \cdot 
\mathbf{f}_m^{\mathrm{D}}(\mathbf{r})\left\{\int_{V_n} \kappa_n(\mathbf{r}^{\prime}) \nabla^{\prime} \cdot 
\mathbf{f}_n^{\mathrm{D}}(\mathbf{r}^{\prime}) G(\mathbf{r}, \mathbf{r}^{\prime})\,dv^{\prime}\right.\\
&\qquad\qquad\qquad\qquad\qquad\qquad
\left.-(\kappa_n^{+}-\kappa_n^{-}) \int_{S_n} 
G(\mathbf{r}, \mathbf{r}^{\prime})\,ds^{\prime}\right\}\,dv \\
& -\frac{1}{\varepsilon_0} \int_{S_m} \hat{\mathbf{n}}_m(\mathbf{r}) \cdot \mathbf{f}_m^{\mathrm{D}}(\mathbf{r})
\left\{\int_{V_n} \kappa_n(\mathbf{r}^{\prime}) \nabla^{\prime} \cdot \mathbf{f}_n^{\mathrm{D}}(\mathbf{r}^{\prime}) 
G(\mathbf{r}, \mathbf{r}^{\prime})\,dv^{\prime}\right.\\
&\qquad\qquad\qquad\qquad\qquad\qquad
\left.-(\kappa_n^{+}-\kappa_n^{-}) \int_{S_n} 
G(\mathbf{r}, \mathbf{r}^{\prime})\,ds^{\prime}\right\}\,ds
\end{aligned}
\end{equation}
where $\hat{\mathbf{n}}_m(\mathbf{r})$ is the unit normal vector on $S_m$ pointing from $V_m^{-}$ to $V_m^{+}$ and $\hat{\mathbf{n}}_m(\mathbf{r}) \cdot \mathbf{f}_m^{\mathrm{D}}(\mathbf{r})=1$. 

\textbf{Case 3:} $\mathbf{f}_m^{\mathrm{D}}(\mathbf{r})$ is a full SWG function and $\mathbf{f}_n^\mathrm{D}(\mathbf{r})$ is a half SWG function:
\begin{equation}
\label{eq:ZDD_fh}
\begin{aligned}
\{\bar{\bar{Z}}_{\mathrm{DD}}\}_{m n}= & \frac{1}{\varepsilon_0 \varepsilon_{\mathrm{b}, n}^{+}} \int_{V_m} \mathbf{f}_m^{\mathrm{D}}(\mathbf{r}) \cdot \mathbf{f}_n^{\mathrm{D}}(\mathbf{r})\,dv-\omega^2 \mu_0 \kappa_n^{+} \int_{V_m} \mathbf{f}_m^{\mathrm{D}}(\mathbf{r}) \cdot \int_{V_n^{+}} \mathbf{f}_n^{\mathrm{D}}(\mathbf{r}^{\prime}) G(\mathbf{r}, \mathbf{r}^{\prime})\,dv^{\prime}\,dv \\
& +\frac{\kappa_n^{+}}{\varepsilon_0} \int_{V_m} \nabla \cdot \mathbf{f}_m^{\mathrm{D}}(\mathbf{r})\left\{\int_{V_n^{+}} \nabla^{\prime} \cdot \mathbf{f}_n^{\mathrm{D}}(\mathbf{r}^{\prime}) G(\mathbf{r}, \mathbf{r}^{\prime})\,dv^{\prime}-\int_{S_n} G(\mathbf{r}, \mathbf{r}^{\prime})\,d s^{\prime}\right\}\,dv.
\end{aligned}
\end{equation}

\textbf{Case 4:} Both $\mathbf{f}_m^{\mathrm{D}}(\mathbf{r})$ and $\mathbf{f}_n^\mathrm{D}(\mathbf{r})$ are half SWG functions:
\begin{equation}
\label{eq:ZDD_hh}
\begin{aligned}
\{\bar{\bar{Z}}_{\mathrm{DD}}\}_{m n}= & \frac{1}{\varepsilon_0 \varepsilon_{\mathrm{b}, n}^{+}} \int_{V_m^{+}} \mathbf{f}_m^{\mathrm{D}}(\mathbf{r}) \cdot \mathbf{f}_n^{\mathrm{D}}(\mathbf{r})\,dv \\
& -\omega^2 \mu_0 \kappa_n^{+} \int_{V_m^{+}} \mathbf{f}_m^{\mathrm{D}}(\mathbf{r}) \cdot \int_{V_n^{+}} \mathbf{f}_n^{\mathrm{D}}(\mathbf{r}^{\prime}) G(\mathbf{r}, \mathbf{r}^{\prime})\,dv^{\prime}\,dv \\
& +\frac{\kappa_n^{+}}{\varepsilon_0} \int_{V_m^{+}} \nabla \cdot \mathbf{f}_m^{\mathrm{D}}(\mathbf{r})\left\{\int_{V_n^{+}} \nabla^{\prime} \cdot \mathbf{f}_n^{\mathrm{D}}(\mathbf{r}^{\prime}) G(\mathbf{r}, \mathbf{r}^{\prime})\,dv^{\prime}-\int_{S_n} G(\mathbf{r}, \mathbf{r}^{\prime})\,ds^{\prime}\right\}\,dv \\
& -\frac{\kappa_n^{+}}{\varepsilon_0} \int_{S_m} \hat{\mathbf{n}}_m(\mathbf{r}) \cdot \mathbf{f}_m^{\mathrm{D}}(\mathbf{r})\left\{\int_{V_n^{+}} \nabla^{\prime} \cdot \mathbf{f}_n^{\mathrm{D}}(\mathbf{r}^{\prime}) G(\mathbf{r}, \mathbf{r}^{\prime})\,dv^{\prime}-\int_{S_n} G(\mathbf{r}, \mathbf{r}^{\prime})\,ds^{\prime}\right\}\,ds.
\end{aligned}
\end{equation}

\subsection{Entries of $\bar{\bar{Z}}_{\mathrm{Dq}}$}
\label{sec:ZDq}
Two possible combinations of the testing function $\mathbf{f}_m^{\mathrm{D}}(\mathbf{r})$ and the basis function $\mathbf{f}_n^{\mathrm{q}}(\mathbf{r})$ must be considered, each leading to a different expression for the matrix entries. Note that $\mathbf{f}_n^{\mathrm{q}}(\mathbf{r})$ is always a full SWG function since the normal components of $\mathbf{J}_{\mathrm{e}}(\mathbf{r})$ and $\mathbf{J}_{\mathrm{h}}(\mathbf{r})$ vanish on $S$.

\textbf{Case 1:} $\mathbf{f}_m^{\mathrm{D}}(\mathbf{r})$ is a full SWG 
function:
\begin{equation}
\label{eq:ZDq_f}
\begin{aligned}
\{\bar{\bar{Z}}_{\mathrm{Dq}}\}_{mn} = & -\frac{1}{\mathrm{j}\omega\varepsilon_0} 
\int_{V_m} \frac{\mathbf{f}_m^{\mathrm{D}}(\mathbf{r}) \cdot 
\mathbf{f}_n^{\mathrm{q}}(\mathbf{r})}{\varepsilon_{\mathrm{b},n}
(\mathbf{r})}\,dv \\
& + \mathrm{j}\omega\mu_0 \int_{V_m} \mathbf{f}_m^{\mathrm{D}}(\mathbf{r}) \cdot 
\int_{V_n} \frac{\mathbf{f}_n^{\mathrm{q}}(\mathbf{r}')}
{\varepsilon_{\mathrm{b},n}(\mathbf{r}')} G(\mathbf{r},\mathbf{r}')\,
dv'\,dv \\
& + \frac{1}{\mathrm{j}\omega\varepsilon_0} \int_{V_m} \nabla \cdot 
\mathbf{f}_m^{\mathrm{D}}(\mathbf{r}) \left\{ \int_{V_n} 
\frac{\nabla' \cdot \mathbf{f}_n^{\mathrm{q}}(\mathbf{r}')}
{\varepsilon_{\mathrm{b},n}(\mathbf{r}')} G(\mathbf{r},\mathbf{r}')\,
dv' \right. \\
&\qquad\qquad\qquad\qquad\qquad\qquad
\left. - \left[\frac{1}{\varepsilon_{\mathrm{b},n}^+} - 
\frac{1}{\varepsilon_{\mathrm{b},n}^-}\right] \int_{S_n} 
G(\mathbf{r},\mathbf{r}')\,ds' \right\}\,dv.
\end{aligned}
\end{equation}

\textbf{Case 2:} $\mathbf{f}_m^{\mathrm{D}}(\mathbf{r})$ is a half SWG 
function:
\begin{equation}
\label{eq:ZDq_h}
\begin{aligned}
\{\bar{\bar{Z}}_{\mathrm{Dq}}\}_{mn} = & -\frac{1}{\mathrm{j}\omega\varepsilon_0} 
\int_{V_m^+} \frac{\mathbf{f}_m^{\mathrm{D}}(\mathbf{r}) \cdot 
\mathbf{f}_n^{\mathrm{q}}(\mathbf{r})}{\varepsilon_{\mathrm{b},n}
(\mathbf{r})}\,dv \\
&+ \mathrm{j}\omega\mu_0 \int_{V_m^+} \mathbf{f}_m^{\mathrm{D}}(\mathbf{r}) \cdot 
\int_{V_n} \frac{\mathbf{f}_n^{\mathrm{q}}(\mathbf{r}')}
{\varepsilon_{\mathrm{b},n}(\mathbf{r}')} G(\mathbf{r},\mathbf{r}')\,
dv'\,dv \\
& + \frac{1}{\mathrm{j}\omega\varepsilon_0} \int_{V_m^+} \nabla \cdot 
\mathbf{f}_m^{\mathrm{D}}(\mathbf{r}) \left\{ \int_{V_n} 
\frac{\nabla' \cdot \mathbf{f}_n^{\mathrm{q}}(\mathbf{r}')}
{\varepsilon_{\mathrm{b},n}(\mathbf{r}')} G(\mathbf{r},\mathbf{r}')\,
dv' \right. \\
&\qquad\qquad\qquad\qquad\qquad\qquad
\left. - \left[\frac{1}{\varepsilon_{\mathrm{b},n}^+} - 
\frac{1}{\varepsilon_{\mathrm{b},n}^-}\right] \int_{S_n} 
G(\mathbf{r},\mathbf{r}')\,ds' \right\}\,dv \\
& - \frac{1}{\mathrm{j}\omega\varepsilon_0} \int_{S_m} 
\hat{\mathbf{n}}_m(\mathbf{r}) \cdot \mathbf{f}_m^{\mathrm{D}}
(\mathbf{r}) \left\{ \int_{V_n} \frac{\nabla' \cdot 
\mathbf{f}_n^{\mathrm{q}}(\mathbf{r}')}{\varepsilon_{\mathrm{b},n}
(\mathbf{r}')} G(\mathbf{r},\mathbf{r}')\,dv' \right. \\
&\qquad\qquad\qquad\qquad\qquad\qquad
\left. - \left[\frac{1}{\varepsilon_{\mathrm{b},n}^+} - 
\frac{1}{\varepsilon_{\mathrm{b},n}^-}\right] \int_{S_n} 
G(\mathbf{r},\mathbf{r}')\,ds' \right\}\,ds.
\end{aligned}
\end{equation}

\subsection{Entries of $\bar{\bar{Z}}_{\mathrm{qD}}$}
\label{sec:ZqD}
Two possible combinations of the testing function $\mathbf{f}_m^{\mathrm{q}}(\mathbf{r})$ and the basis function $\mathbf{f}_n^{\mathrm{D}}(\mathbf{r})$ must be considered, each leading to a different expression for the matrix entries. Note that $\mathbf{f}_m^{\mathrm{q}}(\mathbf{r})$ is always a full SWG function since the normal components of $\mathbf{J}_{\mathrm{e}}(\mathbf{r})$ and $\mathbf{J}_{\mathrm{h}}(\mathbf{r})$ vanish on $S$.

\textbf{Case 1:} $\mathbf{f}_n^{\mathrm{D}}(\mathbf{r})$ is a full SWG 
function:
\begin{equation}
\label{eq:ZqD_f}
\{\bar{\bar{Z}}_{\mathrm{qD}}\}_{mn} = \mathrm{j}\omega\omega_{\mathrm{q}}^2 
\int_{V_m} \frac{\mathbf{f}_m^{\mathrm{q}}(\mathbf{r}) \cdot 
\mathbf{f}_n^{\mathrm{D}}(\mathbf{r})}{\varepsilon_{\mathrm{b},n}
(\mathbf{r})}\,dv.
\end{equation}

\textbf{Case 2:} $\mathbf{f}_n^{\mathrm{D}}(\mathbf{r})$ is a half SWG 
function:
\begin{equation}
\label{eq:ZqD_h}
\{\bar{\bar{Z}}_{\mathrm{qD}}\}_{mn} = \frac{\mathrm{j}\omega\omega_{\mathrm{q}}^2}
{\varepsilon_{\mathrm{b},n}^+} \int_{V_m} 
\mathbf{f}_m^{\mathrm{q}}(\mathbf{r}) \cdot 
\mathbf{f}_n^{\mathrm{D}}(\mathbf{r})\,dv.
\end{equation}

\subsection{Entries of $\bar{\bar{Z}}_{\mathrm{qq}}$}
\label{sec:Zqq}
Since both $\mathbf{f}_m^{\mathrm{q}}(\mathbf{r})$ and $\mathbf{f}_n^{\mathrm{q}}(\mathbf{r})$ are always full SWG functions, only one expression for the matrix entries is needed:
\begin{equation}
\label{eq:Zqq_exp}
\begin{aligned}
\{\bar{\bar{Z}}_{\mathrm{qq}}\}_{mn} = & -\beta_{\mathrm{q}}^2 
\int_{V_m} [\nabla \cdot \mathbf{f}_m^{\mathrm{q}}(\mathbf{r})] [\nabla \cdot \mathbf{f}_n^{\mathrm{q}}(\mathbf{r})]
\,dv + \omega(\omega - \mathrm{j}\gamma_{\mathrm{q}}) \int_{V_m} 
\mathbf{f}_m^{\mathrm{q}}(\mathbf{r}) \cdot 
\mathbf{f}_n^{\mathrm{q}}(\mathbf{r})\,dv \\
& - \omega_{\mathrm{q}}^2 \int_{V_m} 
\frac{\mathbf{f}_m^{\mathrm{q}}(\mathbf{r}) \cdot 
\mathbf{f}_n^{\mathrm{q}}(\mathbf{r})}{\varepsilon_{\mathrm{b},n}
(\mathbf{r})}\,dv.
\end{aligned}
\end{equation}

\subsection{Entries of $\bar{\bar{Z}}_{\mathrm{qp}}$}
\label{sec:Zqp}
Since both $\mathbf{f}_m^{\mathrm{q}}(\mathbf{r})$ and $\mathbf{f}_n^{\mathrm{p}}(\mathbf{r})$ are always full SWG functions, only one expression for the matrix entries is needed:
\begin{equation}
\label{eq:Zqp_exp}
\{\bar{\bar{Z}}_{\mathrm{qp}}\}_{mn} = -\omega_{\mathrm{q}}^2 
\int_{V_m} \frac{\mathbf{f}_m^{\mathrm{q}}(\mathbf{r}) \cdot 
\mathbf{f}_n^{\mathrm{p}}(\mathbf{r})}{\varepsilon_{\mathrm{b},n}
(\mathbf{r})}\,dv.
\end{equation}

\subsection{Entries of $\bar{V}^{\mathrm{inc}}$}
\label{sec:Vinc}
Two possible combinations of the testing function $\mathbf{f}_m^{\mathrm{D}}(\mathbf{r})$ must be considered, each leading to a different expression for the excitation vector entries.

\textbf{Case 1:} $\mathbf{f}_m^{\mathrm{D}}(\mathbf{r})$ is a full SWG 
function:
\begin{equation}
\label{eq:Vinc_f}
\{\bar{V}^{\mathrm{inc}}\}_m = \int_{V_m} \mathbf{f}_m^{\mathrm{D}}
(\mathbf{r}) \cdot \mathbf{E}^{\mathrm{inc}}(\mathbf{r})\,dv.
\end{equation}

\textbf{Case 2:} $\mathbf{f}_m^{\mathrm{D}}(\mathbf{r})$ is a half SWG 
function:
\begin{equation}
\label{eq:Vinc_h}
\{\bar{V}^{\mathrm{inc}}\}_m = \int_{V_m^+} \mathbf{f}_m^{\mathrm{D}}
(\mathbf{r}) \cdot \mathbf{E}^{\mathrm{inc}}(\mathbf{r})\,dv.
\end{equation}

\bibliographystyle{IEEEtran}
\bibliography{references_final}

\newpage\clearpage

\section*{Figures}

\newpage\clearpage

\newpage\clearpage

\newpage\clearpage

%%%%%%%%%% If preparing manually:
% \newcommand{\enquote}[1]{``#1''}

% \bibitem{Zhang:14}
% Y.~Zhang, S.~Qiao, L.~Sun, Q.~W. Shi, W.~Huang, L.~Li, and Z.~Yang,
%   \enquote{Photoinduced active terahertz metamaterials with nanostructured
%   vanadium dioxide film deposited by sol-gel method,}
%   {\protect\JournalTitle{Optics Express}} \textbf{22}, 11070--11078 (2014).

% {Optical Society}, \enquote{{OSA Publishing},}
%   \url{http://www.osapublishing.org}.

% \bibitem{FORSTER2007}
% P.~Forster, V.~Ramaswamy, P.~Artaxo, T.~Bernsten, R.~Betts, D.~Fahey,
%   J.~Haywood, J.~Lean, D.~Lowe, G.~Myhre, J.~Nganga, R.~Prinn, G.~Raga,
%   M.~Schulz, and R.~V. Dorland, \enquote{Changes in atmospheric consituents and
%   in radiative forcing,} in \enquote{Climate Change 2007: The Physical Science
%   Basis. Contribution of Working Group 1 to the Fourth assesment report of
%   Intergovernmental Panel on Climate Change,}  S.~Solomon, D.~Qin, M.~Manning,
%   Z.~Chen, M.~Marquis, K.~B. Averyt, M.~Tignor, and H.~L. Miler, eds.
%   (Cambridge University Press, 2007).

% \end{thebibliography}

\end{document}